\documentclass[11pt]{article}
\usepackage{a4wide}
\usepackage{mathrsfs}
\usepackage{latexsym,bm}
\usepackage{graphicx}
\usepackage{indentfirst}
\usepackage{slashed}
\usepackage{amsmath}
\usepackage{amssymb}
\usepackage[usenames,dvipsnames]{color}
\usepackage{hyperref}
\usepackage{epsfig}
\usepackage[titletoc]{appendix}
\usepackage{multirow}%
\usepackage{rotating}
\usepackage[sort&compress,square,numbers]{natbib}
\usepackage{ulem} %% \sout
\usepackage[top=2.9cm,bottom=2.5cm,left=2.8cm,right=3cm]{geometry}
\usepackage{tabularx}
\usepackage{placeins}
\renewcommand{\arraystretch}{1.2}%for the array in the table

\newcommand{\nn}{\nonumber}

\newcommand{\email}[1]{\footnote{{\em } \texttt{#1}}}

\newcommand{\bma}{\left(\begin{matrix}}
\newcommand{\ema}{\end{matrix}\right)}

\newcommand{\mL}{\mathcal{L}}

\newcommand{\bra}{\langle}
\newcommand{\ket}{\rangle}
\newcommand{\xpt}{{\chi}{\rm PT}}
\newcommand{\rxt}{{\rm R}{\chi}{\rm T}}
\newcommand{\kkbar}{{K\bar{K}}}

\begin{document}
\title{\Large \bf Unified study of scalar, vector and tensor two-meson form factors in $U(3)$ resonance chiral theory}
\author{\small Jin Hao$^a$,\, Chun-Gui Duan$^a$\email{duancg@hebtu.edu.cn},\,  Zhi-Hui Guo$^{a}$\email{zhguo@hebtu.edu.cn},\, J.~Oller$^{b}$\email{oller@um.es},\, J.~Ruiz de Elvira$^{c}$\email{jacobore@ucm.es} \\[0.5em]
{ \small\it ${}^a$ Department of Physics and Hebei Key Laboratory of Photophysics Research and Application, } \\ 
{\small\it Hebei Normal University,  Shijiazhuang 050024, China}\\
{ \small\it ${}^b$ Departamento de F\'{\i}sica, Universidad de Murcia, E-30071 Murcia, Spain}
\\
{ \small\it ${}^c$ Departamento de F\'{\i}sica T\'eorica and IPARCOS, Universidad Complutense de Madrid,} \\
{\small\it E-28040 Madrid, Spain}
}
\date{}

\maketitle

\begin{abstract}
We perform a systematic study of two-meson form factors of the scalar, vector, and anti-symmetric tensor types within the framework of the $U(3)$ resonance chiral theory. The complete perturbative form factors in both the strangeness-conserving and strangeness-changing channels are calculated by incorporating one-loop light-flavor pseudoscalar meson contributions and tree-level resonance exchanges. With these newly calculated chiral results, we construct the corresponding unitarized form factors by incorporating meson-meson final-state interactions. The parameter values obtained in previous meson-meson scattering studies are then exploited to predict the corresponding form factors. Different types of form factors are found to exhibit rather distinct resonance structures across channels. 
\end{abstract} 

\section{Introduction} 

Light-flavor pseudoscalar meson form factors (FFs), namely the meson matrix elements of the quark currents, are critical objects in particle physics. Not only can they provide a bridge to probe the properties of electromagnetic and weak currents in hadron physics, but they are also ideal objects to investigate the non-perturbative QCD dynamics. In this work, we will focus on evaluating the two-meson FFs. 
The pioneer study of the latter quantities~\cite{Gasser:1984ux} in chiral perturbation theory ($\chi$PT) dates back almost to the same time as the establishment of $\chi$PT~\cite{Weinberg:1978kz,Gasser:1983yg,Gasser:1984gg}. The calculations of the scalar and vector two-meson FFs have been pursued up to the two-loop order in the $SU(2)$~\cite{Gasser:1990bv,Bijnens:1998fm}, and $SU(3)$~\cite{Bijnens:2002hp,Bijnens:2003xg} $\chi$PT frameworks. Higher-order $\chi$PT calculation is of great use to precisely pin down the low-energy behavior of the FFs, especially their values and slopes at zero recoiled energy. However, in many phenomenological studies, such as semileptonic and weak decays of the charmed/bottomed hadrons, the hadronic decays of the $\tau$ lepton, the contributions from the hadronic vacuum polarization and the light-by-light processes to the muon $g-2$, and so on, one would also need to explicitly incorporate the hadron resonance contributions in the intermediate energy ranges, where $\chi$PT alone becomes insufficient, since it is based on a low-energy expansion. 

In this regard, resonance chiral theory ($\rxt$)~\cite{Ecker:1988te}, that constructs the interactions between hadronic resonances and light pseudoscalar mesons respecting chiral symmetry, is a valuable approach to extend the $\xpt$ FFs to higher-energy regions. 
Furthermore, the synthesis of chiral amplitudes, unitarity, and analyticity offers a remarkable framework to systematically investigate light-flavor chiral dynamics of the non-perturbative strong interactions both in the low and intermediate energy regions, especially when hadron resonances emerge in the system under study~\cite{Pelaez:2015qba,Oller:2019opk,Yao:2020bxx,Pelaez:2021dak}. 
For the elastic two-meson FF, the Omn\`es function~\cite{Omnes:1958hv} gives a strict solution that fully respects unitarity, analyticity, and Watson final-state theorem~\cite{Watson:1952ji}.  Nevertheless, when extending to the coupled-channel case, it is challenging to set up a strictly rigorous formalism for multichannel FFs, see e.g., recent progress in Refs.~\cite{Shi:2020rkz,VonDetten:2021rax,Kirk:2024oyl,Balz:2025auk}.
In this work, rather than pursuing a purely dispersive analysis, we employ the chiral unitarized approach~\cite{Oller:1997yg,Oller:1998ia} to carry out a comprehensive study of the two-meson FFs with scalar, vector, and antisymmetric tensor (hereafter simply referred to as tensor) structures within the framework of $U(3)$ resonance chiral theory.
Compared to dispersive methods, the chiral unitarized approach offers several important advantages. In particular, since it incorporates the quark-mass dependence dictated by chiral symmetry, it provides a natural and reliable framework for performing chiral extrapolations of lattice QCD results, which also allows one to make correlated predictions for related channels within the same theoretical setup.
Furthermore, within the chiral unitarized approach, the non-perturbative final-state strong interactions of the meson pairs in the FFs are governed by the same objects that also appear in the meson-meson scattering amplitudes. As a consequence, the unitarized FFs can be pure predictions in the chiral unitary approach, since almost all the free parameters in the FFs can be fixed by fitting the scattering amplitudes to the experimental scattering data and relevant event distributions~\cite{Guo:2012ym,Guo:2012yt}. 

The vector FFs of pion and kaon are calculated in Ref.~\cite{Oller:2000ug} within the unitarized $SU(3)$ $\rxt$ by including the octet pseudoscalar $\pi,\,K,\,\eta_8$ and explicit vector resonance states. The calculation of the pion vector FF is further pursued by including the full $1/N_C$ corrections in $\rxt$ in Refs.~\cite{Rosell:2004mn,Pich:2010sm}. 
The strangeness changing scalar FFs of $K\pi$, $K\eta$, and $K\eta'$ are computed in Ref.~\cite{Jamin:2001zq} by combining the tree-level $\rxt$ amplitudes and dispersion relations. In addition, the unitarized FFs are also constructed by taking the $SU(2)$~\cite{Nieves:1999bx} and $SU(3)$~\cite{Shi:2020rkz} $\xpt$ amplitudes with local higher order terms accompanied by low energy constants (LECs). The $U(3)$ chiral framework explicitly incorporates the QCD $U_A(1)$ anomaly effect, which is the main source of the large mass of the singlet $\eta_0$. It has been exploited to calculate the strangeness-conserving scalar FFs in Refs.~\cite{Guo:2012ym,Guo:2012yt} by including the light pseudoscalar meson one-loop amplitudes within $\rxt$. In this work, we push forward the computation of both the strangeness-conserving and changing FFs of the scalar and vector types by including the full light pseudoscalar meson one-loop amplitudes within the framework of $\rxt$. Meanwhile, the tensor FFs~\cite{Cata:2007ns} of the light pseudoscalar mesons are found to play vital roles in searching for the new physics phenomenons beyond the Standard Model (BSM)~\cite{Cirigliano:2017tqn,Miranda:2018cpf,Aguilar:2024ybr,Chen:2019vbr,Chen:2021udz,Hoferichter:2018zwu}. We will calculate these interesting objects by completing two-meson tensor FFs up to the one-loop level in $U(3)$ $\rxt$ and also performing the unitarization. 

This work is organized as follows. In Sec.~\ref{sec.theo}, we first introduce the relevant chiral Lagrangians within the framework of $U(3)$ $\rxt$. Then, by taking into account the light-flavor pseudoscalar meson one-loop contributions and the tree-level resonance-exchange contributions, we calculate the complete perturbative amplitudes of the scalar, vector, and tensor form factors for two-meson systems, and further construct their unitarized counterparts by incorporating the non-perturbative meson-meson final-state interactions. In Sec.~\ref{sec.pheno}, we discuss the phenomenological results of the various scalar, vector, and tensor FFs, with particular attention paid to the resonance structures in different channels. Finally, our conclusions and a brief outlook are given in Sec.~\ref{sec.sum}.

\section{Theoretical framework}\label{sec.theo}

\subsection{Relevant chiral Lagrangian}

We aim at the computation of the full sets of the two-meson FFs with the scalar, vector, and tensor types within the framework of R$\chi$T developed in Ref.~\cite{Ecker:1988te}. Comparing with conventional $\chi$PT~\cite{Gasser:1984gg}, whose dynamical fields are the pseudo-Nambu-Goldstone bosons (pNGBs) $\pi, K, \eta_8$, the heavier degrees of freedom, i.e., the bare/preexisting resonance states, are explicitly incorporated in a chiral covariant way within the framework of R$\chi$T. One could predict the higher-order $\chi$PT LECs in terms of the resonance couplings and masses once the preexisting resonance states are integrated out. The predictions of the $O(p^4)$ LECs from R$\chi$T are found to be in nice agreement with their phenomenological values~\cite{Ecker:1988te}. On the one hand, this indicates that any amplitude calculated within the R$\chi$T framework will automatically respect the chiral symmetry in the low-energy region. On the other hand, the explicit introduction of the heavier preexisting resonance states also provides an opportunity to extend the application region of $\chi$PT with pure pNGBs to higher energy ranges, especially when the resonances start to contribute. Additionally, it is often necessary to combine R$\chi$T with a unitarization approach. This is a must in order to fit data to reaction probabilities (e.g., cross sections or event distributions) as a function of energy. Furthermore, for some channels, the unitarity corrections to the chiral perturbation theory series can be even more important than those stemming from the exchange of preexisting resonances. In particular, they may become so large that they can generate resonances dynamically, like the lightest scalar resonances~\cite{Oller:1997ti,Oller:1998zr}, or the $\Lambda(1405)$~\cite{Oller:2000fj}. Indeed unitarized R$\chi$T has been extensively applied to a wide range of phenomenological processes, such as the meson-meson scattering~\cite{Oller:1998zr,Jamin:2000wn,Guo:2011pa,Nieves:2011gb,Ledwig:2014cla}, two-photon annihilation into two mesons~\cite{Mao:2009cc,Dai:2014zta}, hadronic tau decays~\cite{Arteaga:2022xxy,Gonzalez-Solis:2019lze,Escribano:2016ntp,Escribano:2013bca,Sanz-Cillero:2017fvr,Jamin:2006tk,Guerrero:1997ku,Chen:2022nxm,Guo:2008sh,Guo:2010dv,Li:2025zus,Hao:2025pai}, electron-positron annihilation into multimeson states~\cite{Dai:2013joa,Wang:2023njt,Qin:2024ulb}, etc. 

In this work, not only the preexisting resonance states will be explicitly included, but we also take into account the dynamical field of the singlet $\eta_0$ state, which mixes with the octet $\eta_8$ to give the physical states $\eta$ and $\eta'$. The large mass of the singlet $\eta_0$ component---which dominantly constitutes the physical $\eta'$ and lies around 1~GeV---is widely understood as a consequence of the strong $U_A(1)$ anomaly in QCD. Therefore, the inclusion of the dynamical $\eta_0$ allows us to simultaneously study the processes involving $\pi, K, \eta$ and $\eta'$. The QCD $U_A(1)$ anomaly effect can be systematically incorporated in the chiral framework via the $U(3)$ chiral theory (also dubbed as large $N_C$ chiral theory)~\cite{Herrera-Siklody:1996tqr,Kaiser:2000gs}, which needs to implement the $\delta$ power counting scheme by treating the $1/N_C$ expansion in the same way as the momentum squared and light-flavor quark masses, namely $O(\delta)\sim O(p^2) \sim O(m_q) \sim O(1/N_C)$. The reason behind this is that in the large $N_C$ limit, the QCD $U_A(1)$ anomaly would disappear and the singlet $\eta_0$ would become the ninth pNGB in the chiral limit. As a result, the octet $\pi, K, \eta_8$ and the singlet $\eta_0$ would form the pNGB nonet. 

The leading order (LO) Lagrangian in $U(3)$ chiral theory is given by \cite{Witten:1979vv,Coleman:1980mx,DiVecchia:1980yfw,Kawarabayashi:1980dp}
\begin{eqnarray}\label{eq.lag1}
 \mL_{0}=  \frac{F^2}{4}\bra u_\mu u^\mu \ket+
\frac{F^2}{4}\bra \chi_+ \ket + \frac{F^2}{3} M_0^2 \ln^2{\det u} \,,
\end{eqnarray}
where the chiral tensors are 
\begin{eqnarray}\label{eq.efbb}
&& U =  u^2 = e^{i\frac{ \sqrt2\Phi}{ F}}\,, \qquad \chi = 2 B (s + i p) \,,\qquad \chi_\pm  = u^\dagger  \chi u^\dagger  \pm  u \chi^\dagger  u \,, \nn\\
&& u_\mu = i u^\dagger  D_\mu U u^\dagger \,, \qquad  D_\mu U \, =\, \partial_\mu U - i (v_\mu + a_\mu) U\, + i U  (v_\mu - a_\mu) \,,
\end{eqnarray}
and the matrix of the nonet pNGB fields reads
\begin{equation}\label{eq.phi9}
\Phi \,=\, \left( \begin{array}{ccc}
\frac{1}{\sqrt{2}} \pi^0+\frac{1}{\sqrt{6}}\eta_8+\frac{1}{\sqrt{3}} \eta_0 & \pi^+ & K^+ \\ \pi^- &
\frac{-1}{\sqrt{2}} \pi^0+\frac{1}{\sqrt{6}}\eta_8+\frac{1}{\sqrt{3}} \eta_0   & K^0 \\  K^- & \overline{K}^0 &
\frac{-2}{\sqrt{6}}\eta_8+\frac{1}{\sqrt{3}} \eta_0
\end{array} \right)\,.
\end{equation}

The kinetic Lagrangian of the vector resonance nonet reads~\cite{Ecker:1988te}
\begin{align}\label{eq.kinerv}
\mathcal{L}_{\rm kin}^V &= -\frac{1}{2} \bra \nabla^\lambda
V_{\lambda\mu}\nabla_\nu V^{\nu\mu}
-\frac{1}{2}M^2_V V_{\mu\nu}V^{\mu\nu} \ket \,, 
\end{align}
where the anti-symmetric tensor is adopted to describe the vector resonances. 
For the scalar ($S$) and pseudoscalar ($P$) resonances, their kinetic Lagrangians are given by  
\begin{align}\label{eq.kinersp}
\mathcal{L}_{\rm kin}^{R} &= \frac{1}{2} \bra \nabla^\mu R_8 \nabla_\mu R_8
-M^2_{R_8} R_8^{2} \ket +\frac{1}{2}\big(\partial^\mu R_{1}
\partial_\mu R_{1}-M^2_{R_1} R_1^{2}\big)~,
\end{align}
where $R_8$ and $R_1$ with $R=(S,P)$ denote the $SU(3)$ octet and singlet, respectively.  
The minimal Lagrangian describing the interaction of vector resonances consists of two operators~\cite{Ecker:1988te}
\begin{eqnarray}\label{eq.lagvector}
\mL_{2,V}= \frac{F_V}{2\sqrt{2}} \langle V_{\mu\nu} f_+^{\mu\nu} \rangle + \frac{iG_V}{\sqrt{2}} \langle V_{\mu\nu} u^\mu u^\nu \rangle\,. 
\end{eqnarray}
For the scalar and pseudoscalar resonances, we follow Refs.~\cite{Guo:2012ym,Guo:2012yt} to separately treat the interaction operators involving the octet and singlet states
\begin{align}\label{eq.lagscalar}
\mL_{S}&= c_d\bra S_8 u_\mu u^\mu \ket + c_m \bra S_8 \chi_+ \ket
 + \tilde{c}_d S_1 \bra u_\mu u^\mu \ket
+ \tilde{c}_m  S_1 \bra  \chi_+ \ket \,,\\ 
\mL_{P} &= i d_m \bra P_8 \chi_- \ket + i \tilde{d}_m  P_1 \bra \chi_- \ket\,. \label{eq.lagpscalar}
\end{align}
The operators in Eq.~\eqref{eq.lagpscalar} will cause mixing terms between the pseudoscalar resonances and pNGBs, which can be eliminated by performing proper field redefinitions. See Ref.~\cite{Guo:2012yt} for details about this procedure, and we do not repeat it here.  

Compared to the vector and scalar sectors, the phenomenology of pseudoscalar resonances remains considerably less understood. This is mainly due to the scarcity and limited precision of experimental data, the large widths and strong overlapping among states, and the complications associated with their mixing with the pNGBs and with the nontrivial effects of the $U_A(1)$ anomaly. Consequently, the pseudoscalar resonance parameters are affected by sizable uncertainties. 
To partially account for these uncertainties, an $L_8$-like operator is introduced in Ref.~\cite{Guo:2012ym}
\begin{eqnarray} \label{eq.deltaL8}
\frac{\delta L_8}{2} \langle \chi_+ \chi_+ + \chi_- \chi_- \rangle.
\end{eqnarray}
It is noted that $\delta L_8$ is different from $L_8$ in $\xpt$ and their relation can be written as $L_8^{\chi PT}=L_8^{\rm Res}+\delta L_8$, where $L_8^{\rm Res}$ corresponds to the resonance contributions at leading $N_C$ and $\delta L_8$ can be interpreted as the remnant part sub-leading in $1/N_C$. Finally, two additional terms in the $U(3)$ Lagrangian are also considered 
\begin{eqnarray}\label{eq.laglam12}
\mL_\Lambda=\Lambda_1\frac{F^2}{12} D_\mu\psi D^\mu\psi-i\Lambda_2\frac{F^2}{12}\langle U^\dagger\chi- \chi^\dagger U\rangle\psi,
\end{eqnarray}
with 
\begin{eqnarray}
\psi=-i\,\text{ln\,det}U,\quad D_\mu\psi=\partial_\mu \psi-2 \langle a_\mu \rangle, \quad a_\mu=(r_\mu-l_\mu)/2\,. 
\end{eqnarray}
The two operators in Eq.~\eqref{eq.laglam12} are sub-leading in $1/N_C$ and can not be generated by the resonance Lagrangians given above. 

To calculate the tensor FF in chiral effective field theory, it is convenient to introduce the antisymmetric tensor source $\bar{t}^{\mu\nu}$ in the QCD Lagrangian $\bar{q}\sigma_{\mu\nu}\bar{t}^{\mu\nu}q$. The corresponding formalism in $\xpt$ was first studied in Ref.~\cite{Cata:2007ns}.
The LO $\xpt$ Lagrangian coupled to the tensor external source scales as $\mathcal{O}(p^4)$ and the relevant operator to our study is given by 
\begin{equation}\label{eq.lagtc1}
\mL_T^{(4)}=-i\Lambda_2^{T} \langle t_+^{\mu\nu}u_\mu u_\nu \rangle ,
\end{equation}
with 
\begin{eqnarray}
&t_{\pm}^{\mu\nu} = u^{\dagger} t^{\mu\nu} u^{\dagger} \pm u \, t^{\mu\nu\dagger} u \,, \quad t^{\mu \nu} = \frac{1}{4} \big( g^{\mu \lambda} g^{\nu \rho} - g^{\nu \lambda} g^{\mu \rho} - i \epsilon^{\mu \nu \lambda \rho} \big) \bar{t}_{\lambda \rho}\,.
\end{eqnarray}
For the Levi-Civita tensor $\epsilon^{\mu\nu\lambda\rho}$, we use the convention of $\epsilon^{0123}=1$ throughout. In addition, we also include the resonance contributions to the tensor FFs. 
The corresponding resonance Lagrangian, involving a single resonance field and the lowest-order couplings to the external tensor source, reads~\cite{Miranda:2018cpf}
\begin{equation}\label{eq.lagtv} 
\mL_{T,V} = F_V^T M_V \langle V_{\mu\nu} t^{\mu\nu}_+\rangle \,,
\end{equation}
where the tensor coupling $F_V^{T}$ has mass dimension. Possible terms from the axial-vector resonances are expected to be tiny and hence are neglected.  

\subsection{Calculation of the scalar form factors}

By including tree-level resonance exchange together with the pNGB one-loop diagrams, the strangeness-conserving two-meson scalar FFs were computed in Refs.~\cite{Guo:2012ym,Guo:2012yt}. As one of the main novelties of the present work, we extend this analysis to the strangeness-changing channels within the same framework. In addition, our calculation improves upon Ref.~\cite{Jamin:2001zq}, which only accounted for tree-level resonance exchanges, by incorporating the full set of pNGB one-loop contributions.

The two-meson scalar FF corresponds to the matrix element of the scalar quark current, i.e.,
\begin{equation}
F_{S,PQ}^{mn}(s) = \frac{1}{B}\langle 0| \bar{q}^m q^n |  PQ \rangle\,, 
\end{equation}
where $m$ and $n$ denote the quark flavor indices, $S$ in the subscript refers to the scalar FF,  $P$ and $Q$ stand for the two mesons in the final states and $s$ is the energy squared of the $PQ$ system. The results will be presented using the quark-flavor basis rather than the singlet-octet basis for the quark currents. In this basis, the combinations $(\bar{u}u+\bar{d}d)/\sqrt{2}$ and $\bar{s}s$ constitute the isoscalar sector, while the bilinears $(\bar{u}u-\bar{d}d)/\sqrt{2}$, $\bar{u}d$, and $\bar{d}u$ form an isovector triplet.
For strangeness-changing transitions, the relevant operators arrange into two isospin-$1/2$ doublets, namely $(\bar{d}s,\bar{u}s)$ and $(\bar{s}u,\bar{s}d)$. Since CP-violating effects are not considered in this work, these two doublets are related by charge conjugation and therefore governed by identical strong interaction dynamics.
Furthermore, we work in the isospin ($I$) symmetric limit throughout, implying that there are four independent scalar quark currents, which are chosen as $(\bar{u}u+\bar{d}d)/\sqrt{2} \,(I=0)\,,\bar{s}s \,(I=0)\,, \bar{u}d \,(I=1)\,,  \bar{u}s (I=\frac{1}{2})$. In Table~\ref{tab.sff}, we summarize the two-meson channels coupled to the various scalar currents and find that there are 16 independent two-meson scalar FFs of the four types of currents. In order to construct the two-meson FFs in the isospin basis, we follow the same conventions as those given in Eq.~(19) of Ref.~\cite{Guo:2011pa}, which were employed in the $U(3)$ meson-meson scattering case. For the charged kaon states, however, we use the phase convention
\begin{eqnarray}
|K^+\rangle=\left|\frac12,\frac12\right\rangle\,,\qquad
|K^-\rangle=-\left|\frac12,-\frac12\right\rangle\,,
\end{eqnarray}
in the subsequent analysis of the FFs, in order to be consistent with Refs.~\cite{Oller:1998hw,Shi:2020rkz}. This only amounts to a different phase convention for the charged kaon isospin components and does not affect the remaining results of Ref.~\cite{Guo:2011pa,Guo:2012ym,Guo:2012yt}. The isoscalar two-meson states are given by
\begin{eqnarray}\label{eq.ffisospin0norm}
|\pi\pi \ket_{I=0} &=& - \frac{ |\pi^+ \pi^-\ket + |\pi^- \pi^+ \ket + |\pi^0 \pi^0 \ket}{\sqrt{6}}\,,\nonumber \\
|K\bar{K} \ket_{I=0} &=& - \frac{ |K^+ K^-\ket + |K^0 \bar{K}^0 \ket }{\sqrt{2}}\,,\nonumber \\
|\eta\eta \ket_{I=0} &=&  \frac{ |\eta \eta \ket }{\sqrt{2}}\,,\nonumber \\
|\eta\eta' \ket_{I=0} &=& |\eta \eta' \ket \,,\nonumber \\
|\eta'\eta' \ket_{I=0} &=& \frac{ |\eta' \eta' \ket }{\sqrt{2}}\,,
\end{eqnarray}
and the isovector two-meson states are 
\begin{eqnarray}\label{eq.ffisospin1norm}
|\pi\pi \ket_{I=1} &=&  \frac{|\pi^0\pi^- \ket-|\pi^-\pi^0 \ket}{2} \,,\nonumber \\
|K\bar{K} \ket_{I=1} &=& -|K^0K^- \ket\,,\nonumber \\
|\pi\eta \ket_{I=1} &=&  |\pi^- \eta \ket \,,\nonumber \\
|\pi\eta' \ket_{I=1} &=& |\pi^- \eta' \ket \,,
\end{eqnarray}
where the symmetry factor of $\sqrt{2}$ is introduced for the identical particle pairs, such as the $\pi\pi$, $\eta\eta$, and $\eta'\eta'$, consistent with the convention of the partial-wave meson-meson amplitudes in Refs.~\cite{Guo:2011pa,Guo:2012ym,Guo:2012yt}. The two-meson states in the isospin-1/2 case are defined as 
\begin{eqnarray}\label{eq.ffisospin1d2norm}
|\bar{K}\pi \ket_{I=\frac{1}{2}} &=&  -\frac{\sqrt{2}|\bar{K}^0\pi^-\ket+|K^-\pi^0 \ket}{\sqrt{3}} \,,\nonumber\\ 
|\bar{K}\eta \ket_{I=\frac{1}{2}} &=&  -|K^-\eta \ket \,,\nonumber\\
|\bar{K}\eta' \ket_{I=\frac{1}{2}} &=&  -|K^-\eta'\ket \,. 
\end{eqnarray}

\begin{table}[htbp]
\centering
\begin{tabular}{|c|c|c|}
\hline
Isospin, Strangeness & Scalar currents& Two-meson channels\\
\hline
\multirow{2}{*}{$I=0,S=0$}& $(\bar{u}u+\bar{d}d)/\sqrt{2}$   & $\pi\pi,\,K\bar{K},\,\eta\eta,\,\eta\eta', \,\eta'\eta'$ \\
\cline{2-3}
 & $\bar{s}s$ & $\pi\pi,\,K\bar{K},\,\eta\eta,\,\eta\eta', \,\eta'\eta'$\\
\hline
$I=1,S=0$& $\bar{u}d$&$\pi\eta,\,K\bar{K},\,\pi\eta'$ \\
\hline
$I=\frac{1}{2},S=-1$& $\bar{u}s$ &$\bar{K}\pi \,,\,\bar{K}\eta\,,\, \bar{K}\eta'$\\
\hline
\end{tabular}
\caption{Relevant two-meson channels coupled to the various scalar currents.}
\label{tab.sff}
\end{table}

\begin{figure}[htbp]
    \centering
      \includegraphics[width=0.9\linewidth]{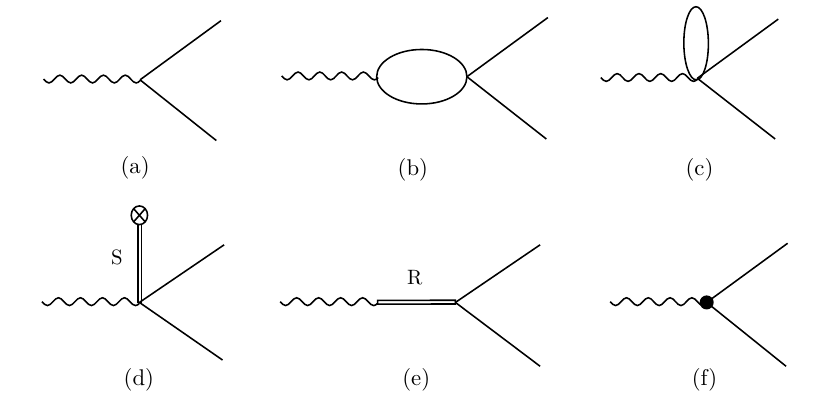}    
    \caption{Feynman diagrams for the two-meson FFs. The wiggly lines denote the scalar/vector/tensor external sources for the scalar/vector/tensor FFs. The double lines denote the proper resonance fields. The diagram (d) only appears in the scalar and vector FFs, and the circle cross symbol connected to the scalar resonance stands for the vacuum, which is caused by the tadpole contribution from the $c_m$ and $\tilde{c}_m$ terms in Eq.~\eqref{eq.lagscalar}.  The type of the resonance $R$ in diagram (e) depends on the different FFs, $R=S$ for scalar and $R=V$ for vector and tensor. The diagram (f) arises from the two local NLO operators in Eq.~\eqref{eq.laglam12}, the pseudoscalar resonance operator in Eq.~\eqref{eq.lagpscalar} after performing the appropriate field redefinitions, and the additional $L_8$-like operator in Eq.~\eqref{eq.deltaL8}, which only enters in the scalar FFs.}
    \label{fig.feyndiag}
\end{figure}

\begin{figure}[htbp]
    \centering
        \includegraphics[width=0.9\linewidth]{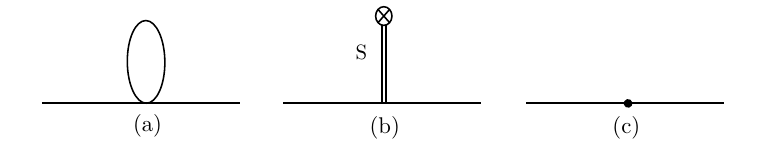}    
    \caption{Feynman diagrams contributing to the pNGB self-energy. The meanings of the different symbols are illustrated in Fig.~\ref{fig.feyndiag}.}
    \label{fig.selfenergy}
\end{figure}

The relevant Feynman diagrams to the FFs calculated in this work are illustrated in Fig.~\ref{fig.feyndiag}, which include both the tree-level resonance exchanges and the pNGB one-loop effects. It should also be noted that the wave function renormalizations of the pNGB fields, as shown in Fig.~\ref{fig.selfenergy}, are also taken into account.  
The LO expressions of the various FFs defined by the $\bar{u}s$ current, corresponding to the diagram (a) of Fig.~\ref{fig.feyndiag}, are 
\begin{eqnarray}
F_{S,K^- \pi^0}^{ \bar{u}s,\mathrm{LO}}= \frac{1}{\sqrt{2}}\,,\quad
F_{S,K^- \eta}^{\bar{u}s,\text{LO}} = -\frac{\sqrt{6} c_\theta + 4\sqrt{3} s_\theta}{6} \,, \quad
F_{S,K^- \eta'}^{  \bar{u}s,\text{LO}} = \frac{4\sqrt{3} c_\theta - \sqrt{6} s_\theta}{6} \,,
\end{eqnarray}
where $c_\theta=\cos\theta$ and $s_\theta=\sin\theta$, and $\theta$ denotes the LO $\eta$-$\eta'$ mixing angle
\begin{eqnarray}
\sin{\theta} &=& -\left( \sqrt{1 +
\frac{ \big(3M_0^2 - 2\Delta^2 +\sqrt{9M_0^4-12 M_0^2 \Delta^2 +36 \Delta^4 } \big)^2}{32 \Delta^4} } ~\right)^{-1}\,,
\label{eq.deftheta0}
\end{eqnarray}
with $\Delta^2 = m_{\overline{K}}^2 - m_{\overline{\pi}}^2$, and $m_{\overline{\pi}}$ and $m_{\overline{K}}$ denoting the LO masses of the pion and kaon, respectively. Although the LO expressions of the scalar FFs with isoscalar and isovector quark currents have been given in Ref.~\cite{Guo:2012yt}, we present the results here for the sake of completeness. 
The LO results for the isoscalar current are 
\begin{equation}
\renewcommand{\arraystretch}{1.6}
\begin{array}{@{}l@{\hspace{0.5 cm}}l@{}}
\displaystyle
F_{S,\pi^0 \pi^0}^{\frac{\bar{u}u+\bar{d}d}{\sqrt2},\mathrm{LO}}
= \sqrt{2},
&
\displaystyle
F_{S,\pi^0 \pi^0}^{\bar{s}s,\mathrm{LO}}
= 0,
\\
\displaystyle
F_{S,K^0 \bar{K}^0}^{\frac{\bar{u}u+\bar{d}d}{\sqrt2},\mathrm{LO}}
= \frac{1}{\sqrt{2}},
&
\displaystyle
F_{S,K^0 \bar{K}^0}^{\bar{s}s,\mathrm{LO}}
= 1,
\\
\displaystyle
F_{S,\eta \eta}^{\frac{\bar{u}u+\bar{d}d}{\sqrt2},\mathrm{LO}}
= -\frac{\sqrt{2}}{3}
\left(c_\theta^2+2\sqrt{2}c_\theta s_\theta-2\right),
&
\displaystyle
F_{S,\eta \eta}^{\bar{s}s,\mathrm{LO}}
= \frac{2}{3}
\left(c_\theta^2+2\sqrt{2}c_\theta s_\theta+1\right),
\\
\displaystyle
F_{S,\eta \eta'}^{\frac{\bar{u}u+\bar{d}d}{\sqrt2},\mathrm{LO}}
= \frac{\sqrt{2}}{3}
\left(\sqrt{2}c_\theta^2-c_\theta s_\theta-\sqrt{2}s_\theta^2\right),
&
\displaystyle
F_{S,\eta \eta'}^{\bar{s}s,\mathrm{LO}}
= -\frac{2}{3}
\left(\sqrt{2}c_\theta^2-c_\theta s_\theta-\sqrt{2}s_\theta^2\right),
\\
\displaystyle
F_{S,\eta' \eta'}^{\frac{\bar{u}u+\bar{d}d}{\sqrt2},\mathrm{LO}}
= \frac{\sqrt{2}}{3}
\left(-s_\theta^2+2\sqrt{2}c_\theta s_\theta+2\right),
&
\displaystyle
F_{S,\eta' \eta'}^{\bar{s}s,\mathrm{LO}}
= \frac{2}{3}
\left(s_\theta^2-2\sqrt{2}c_\theta s_\theta+1\right),
\end{array}
\end{equation}
and the scalar FFs of the isovector current are given by
\begin{eqnarray}
F_{S,\pi^- \eta}^{\bar{u}d,\mathrm{LO}}
= \frac{\sqrt{2}c_\theta-2s_\theta}{\sqrt{3}},\quad\quad\quad
F_{S,K^0 K^-}^{\bar{u}d,\mathrm{LO}}
= 1,\quad\quad\quad
F_{S,\pi^- \eta'}^{\bar{u}d,\mathrm{LO}}
= \frac{2c_\theta+\sqrt{2}s_\theta}{\sqrt{3}}.
\end{eqnarray}
The complete expressions for the strangeness-changing ($I=\frac{1}{2}$) scalar FFs are given in Eqs.~\eqref{SFFbegin}--\eqref{SFFend} of the Appendix. The explicit expressions for the remaining scalar FFs with isoscalar and isovector quark currents have been provided in Ref.~\cite{Guo:2012yt}.

For the scalar FFs, when restricting the intermediate states to two-meson channels, unitarity implies the following relation above the corresponding thresholds:
\begin{equation}\label{eq.imf}
\mathrm{Im}\, F_{S,j}^I(s)=\sum_k T_{jk}^{IJ}(s)^*\,
\rho_k(s)\, F_{S,k}^I(s)\, .
\end{equation}
Here $T_{jk}^{IJ}(s)$ denotes the meson--meson scattering partial-wave amplitude with isospin $I$ and angular momentum $J$ (with $J=0$ for scalar FFs), while $j$ and $k$ label the coupled two-meson channels and $\rho_k(s)$ denotes the kinematical factor  
\begin{equation}
\rho_k(s) = \frac{\sqrt{[s - (m_a + m_b)^2][s - (m_a - m_b)^2]}}{16\pi s} \equiv \frac{q_k(s)}{8\pi\sqrt{s}}\,,
\end{equation}
where $m_a$ and $m_b$ are the masses of the two mesons in the channel $k$, and $q_k(s)$ is the three momentum in the center of mass (CM) frame. However, the perturbative FFs and scattering amplitudes calculated within chiral effective field theory do not fulfill the unitarity relation in Eq.~\eqref{eq.imf} exactly. Instead, unitarity is satisfied only perturbatively, namely order by order in the $\xpt$ expansion.

In order to reliably describe the resonance dynamics, one needs to restore the unitarity relations. For the elastic case when there is only a single channel, one can strictly solve Eq.~\eqref{eq.imf} and write the FF in terms of the Omn\'es function~\cite{Omnes:1958hv}. However, its rigorous extension to coupled channels is considerably more involved, since it requires as input the coupled-channel scattering amplitudes over a broad energy region, where sizable uncertainties are typically present. In addition, the Omn\'es solution requires knowledge of the phase of the FF up to asymptotically high energies, which may introduce further important uncertainties. In this work, instead of relying on the dispersive method, we will follow Refs.~\cite{Guo:2012ym,Guo:2012yt} to implement the unitarization of the scalar FFs using the following formula
\begin{equation}\label{eq.unif}
F_S^I(s) = \left[ 1 + N^{IJ}(s)\cdot g^{IJ}(s) \right]^{-1} \cdot R_S^I(s),   
\end{equation}
with 
\begin{eqnarray}\label{eq.defnij}
N^{IJ}(s) = T^{IJ}(s)^{(2)+\text{Res+Loop}} + T^{IJ}(s)^{(2)}\cdot g^{IJ}(s)\cdot T^{IJ}(s)^{(2)}\,,
\end{eqnarray}
\begin{equation}
R_S^I(s) = F_S^I(s)^{(2)+\text{Res+Loop}} + N^{IJ}(s)^{(2)}\cdot g^{IJ}(s)\cdot F_S^I(s)^{(2)}\,,
\end{equation}
where $J=0$ should be taken for scalar FFs. 
Here, $T^{IJ}(s)^{(2),\mathrm{Res},\mathrm{Loop}}$ and $F_S^{I}(s)^{(2),\mathrm{Res},\mathrm{Loop}}$ denote, respectively, the $\mathcal{O}(p^2)$, tree-level resonance-exchange, and pNGB loop contributions to the perturbative meson--meson scattering amplitudes and two-meson scalar FFs calculated within $\rxt$. Furthermore, $g^{IJ}(s)$ is an $n\times n$ diagonal matrix whose $k$th nonvanishing diagonal entry is given by
\begin{align}
16\pi^2\, g_k^{IJ}(s) &= a_{SL}(\mu) + \log\frac{m_b^2}{\mu^2} 
- x_+ \log\frac{x_+ - 1}{x_+} - x_- \log\frac{x_- - 1}{x_-}, \\
x_{\pm} &= \frac{s + m_a^2 - m_b^2}{2s} 
\pm \frac{1}{2s} \sqrt{-4s(m_a^2 - i0^+) + \left(s + m_a^2 - m_b^2\right)^2}\,,
\end{align}
corresponding to the two-point one-loop function with the subtraction constant $a_{SL}$. The imaginary part of $g_k^{IJ}(s)$ turns out to be ${\rm Im}g_k^{IJ}(s)=-\rho_k(s)$. 
It is then straightforward to demonstrate that $N^{IJ}(s)$ and $R_S(s)$ do not contain right-hand cuts, see Refs.~\cite{Guo:2011pa,Guo:2012ym,Guo:2012yt} for details. The unitarized meson-meson scattering can be similarly constructed as 
\begin{equation}\label{eq.unit}
T^{IJ}(s) = \left[ 1 + N^{IJ}(s)\cdot g^{IJ}(s) \right]^{-1}\cdot N^{IJ}(s)\,.
\end{equation}

The unitarity relation given in Eq.~\eqref{eq.imf} is automatically satisfied by the unitarized FF $F_S^{I}(s)$~\eqref{eq.unif} and meson-meson scattering amplitude $T^{IJ}(s)$~\eqref{eq.unit} above thresholds. It should be noted that, compared to the unitarized scattering amplitude $T^{IJ}(s)$, no additional unknown parameters enter in the unitarized FF $F_S^{I}(s)$. In other words, once the unitarized scattering amplitudes are fixed by the relevant meson-meson scattering data, we can give pure predictions to the unitarized scalar FFs. Both the theoretical formulas and phenomenological results of the strangeness-conserving scalar FFs with $I=0$ have been given in Refs.~\cite{Guo:2012ym,Guo:2012yt}. Here, 
our focus is primarily on evaluating the strangeness-changing scalar FFs. In addition, we also update the strangeness-conserving scalar FFs with $I=1$ by taking the revised parameters from Ref.~\cite{Guo:2016zep}, which are obtained by fitting lattice finite-volume energy levels of the $\pi\eta$ and $K\bar{K}$ system.  

For the scalar FFs of the physical states with isoscalar quark currents, their relations can be easily obtained as
\begin{eqnarray}
F_{S,\pi^0\pi^0}^{\frac{\bar{u}u+\bar{d}d}{\sqrt2}}=F_{S,\pi^+\pi^-}^{\frac{\bar{u}u+\bar{d}d}{\sqrt2}}  = F_{S,\pi^-\pi^+}^{\frac{\bar{u}u+\bar{d}d}{\sqrt2}}\,, \quad 
F_{S,K^+K^-}^{\frac{\bar{u}u+\bar{d}d}{\sqrt2}} = F_{S,K^0\bar{K}^0}^{\frac{\bar{u}u+\bar{d}d}{\sqrt2}}\,, \quad 
\end{eqnarray}
and
\begin{eqnarray}
F_{S,\pi^0\pi^0}^{\bar{s}s} = F_{S,\pi^+\pi^-}^{\bar{s}s} = F_{S,\pi^-\pi^+}^{\bar{s}s}\,, \quad 
F_{S,K^+K^-}^{\bar{s}s} = F_{S,K^0\bar{K}^0}^{\bar{s}s} \,. 
\end{eqnarray}
For the scalar FFs with isovector currents, we have the following relations
\begin{eqnarray}
F_{S,\pi^0 \eta^{(')}}^{\frac{\bar{u}u - \bar{d}d}{\sqrt2}} = F_{S,\pi^- \eta^{(')}}^{\bar{u}d} = F_{S,\pi^+ \eta^{(')}}^{\bar{d}u}\,,\quad
F_{S,K^+ K^-}^{\frac{\bar{u}u - \bar{d}d}{\sqrt2}} = -F_{S,K^0 \bar{K}^0}^{\frac{\bar{u}u - \bar{d}d}{\sqrt2}} = \frac{1}{\sqrt{2}} F_{S,K^0 K^-}^{\bar{u}d} = \frac{1}{\sqrt{2}} F_{S,K^+\bar{K}^0 }^{\bar{d}u}\,.
\end{eqnarray}
For the scalar FFs with isospin 1/2 quark currents, their relations are given by 
\begin{eqnarray}
&F_{S,K^0 \pi^+}^{\bar{s} u} = F_{S,\bar{K}^0 \pi^-}^{\bar{u} s} 
= F_{S,K^+ \pi^-}^{\bar{s} d} = F_{S,K^- \pi^+}^{ \bar{d} s} \nonumber  
= \sqrt{2} F_{S,K^+ \pi^0}^{\bar{s} u} \\
&= \sqrt{2} F_{S,K^- \pi^0}^{\bar{u} s} 
= -\sqrt{2} F_{S,K^0 \pi^0}^{\bar{s} d} = -\sqrt{2} F_{S,\bar{K}^0 \pi^0}^{\bar{d} s}\,, \nonumber \\
&F_{S,K^+ \eta^{(')}}^{\bar{s} u} = F_{S,K^- \eta^{(')}}^{\bar{u} s} 
= F_{S,K^0 \eta^{(')}}^{\bar{s} d} = F_{S,\bar{K}^0 \eta^{(')}}^{\bar{d} s}\,.
\end{eqnarray}

\subsection{Calculation of the vector form factors}

In this part, we calculate the vector FF, and it can be extracted from the matrix element of the vector current via 
\begin{equation} \label{260503.1}
\langle 0| \bar{q}^m \gamma^\mu q^n |P(p_1)Q(p_2)\rangle \equiv (p_2-p_1)^\mu F^{mn}_{V+,PQ}(s) + (p_2+p_1)^\mu F^{mn}_{V-,PQ}(s)\,, 
\end{equation}
where $F^{mn}_{V+,PQ}(s)$ corresponds to the vector FF (hereinafter abbreviated as $F^{mn}_{+,PQ}(s)$ in this work) and $F^{mn}_{V-,PQ}(s)$ is the scalar component of matrix element. The scalar FFs discussed previously can be related to the quantities of $F^{mn}_{V+,PQ}(s)$ and $F^{mn}_{V-,PQ}(s)$~\cite{Gasser:1984ux}. 
After taking into account the charge conjugate and isospin symmetries, there are seven types of independent two-meson vector FFs, as illustrated in Table~\ref{tab.vff}. 
The perturbative vector FFs in $\rxt$ can be obtained by calculating the corresponding Feynman diagrams in Fig.~\ref{fig.feyndiag}, where wavy lines should be taken as the vector external sources. 
The explicit LO expressions of the seven independent vector FFs read  
\begin{eqnarray}
&F^{\frac{\bar{u}u+\bar{d}d}{\sqrt2},\text{LO}}_{+,K^+K^-}=-\frac{1}{\sqrt{2}},\quad F^{\bar{s}s,\text{LO}}_{+,K^+K^-}=1 ,\quad   F^{\bar{u}d,\text{LO}}_{+,\pi^-\pi^0}=-\sqrt{2},\quad F^{\bar{u}d,\text{LO}}_{+,K^0K^-}=-1 ,\nonumber \\ 
& F^{\bar{u}s,\text{LO}}_{+,K^-\pi^0}=-\frac{1}{\sqrt{2}},\quad F^{\bar{u}s,\text{LO}}_{+,K^-\eta}=-\frac{\sqrt{6}}{2}c_{\theta},\quad F^{\bar{u}s,\text{LO}}_{+,K^-\eta'}=-\frac{\sqrt{6}}{2}s_{\theta}\,.
\end{eqnarray}
Due to the lengthy formulas, the complete expressions for these seven perturbative vector FFs are not displayed in the main text, but, instead, presented in Eqs.~\eqref{VFFbegin}--\eqref{VFFend} of the Appendix.

\begin{table}
\centering
\begin{tabular}{|c|c|c|}
\hline
 Isospin, Strangeness & Vector currents& Two-meson channels\\
\hline
\multirow{2}{*}{$I=0,S=0$}& $\frac{\bar{u}\gamma^\mu u+\bar{d}\gamma^\mu d}{\sqrt{2}}$   & $K\bar{K}$ \\
\cline{2-3}
& $\bar{s}\gamma^\mu s$  & $K\bar{K}$ \\
\hline
$I=1,S=0$& $\bar{u}\gamma^\mu d$&$ \pi\pi \,,\,K\bar{K}$ \\
\hline
$I=\frac{1}{2},S=-1$& $\bar{u}\gamma^\mu s$ &$\bar{K}\pi \,,\,\bar{K}\eta\,,\, \bar{K}\eta'$\\
\hline
\end{tabular}
\caption{Relevant two-meson channels coupled to vector currents.}
\label{tab.vff}
\end{table}

Next, we perform the unitarization of the vector FF $F^{I}_{+}(s)$ with definite isospin $I$, whose unitarity relation can be expressed in the matrix form as~\cite{Oller:2000ug}
\begin{equation}\label{eq.imfv}
\text{Im}\, F^I_+(s) = \tilde{Q}(s)^{-1} \cdot T^{IJ}(s)^* \cdot \rho(s)  \cdot\tilde{Q}(s) \cdot F^{I}_+(s)\,,
\end{equation}
where the diagonal matrix $\tilde{Q}$ is defined by $\tilde{Q}_{ij}(s)=q_j(s)\delta_{ij}$ and $J=1$ should be taken for vector FFs.  
Following the unitarization procedure presented in the previous subsection, we build the unitarized vector FFs 
\begin{equation}\label{eq.unifv}
F^I_+(s) = \tilde{Q}(s)^{-1} \cdot \left[ 1 + N^{IJ}(s) \cdot g^{IJ}(s) \right]^{-1} \cdot \tilde{Q}(s) \cdot R_+^I(s)\,,
\end{equation}
\begin{equation}
R_+^I(s) = F_+^I(s)^{(2)+\text{Res+Loop}} + \tilde{Q}(s)^{-1} \cdot N^{IJ}(s)^{(2)}\cdot \, g^{IJ}(s)\cdot \,\tilde{Q}(s) \cdot F_+^I(s)^{(2)}\,,
\end{equation}
where $N^{IJ}$ has been given in Eq.~\eqref{eq.defnij} and the superscripts (2), Res, Loop stand for the contributions from the LO, tree-level resonance exchange and pNGB one-loop diagrams, respectively, to the vector FFs in $\rxt$. It is straightforward to verify that the unitarized vector FF of Eq.~\eqref{eq.unifv} and the unitarized meson-meson scattering amplitude of Eq.~\eqref{eq.unit} satisfy the unitarity relation shown in Eq.~\eqref{eq.imfv}. 

By using the seven independent types of unitarized vector FFs in the isospin bases as given in Table~\ref{tab.vff}, one can straightforwardly compute the two-meson vector FFs of different physical states. The relations of the vector FFs with isoscalar quark currents for different physical states are 
\begin{eqnarray}\label{eq.vff0relt}
&F^{\frac{\bar{u}u + \bar{d}d}{\sqrt2}}_{+,K^+K^-}=F^{\frac{\bar{u}u + \bar{d}d}{\sqrt2}}_{+,K^0\bar{K}^0}, \quad F^{\bar{s}s}_{+,K^+K^-}=F^{\bar{s}s}_{+,K^0\bar{K}^0}.
\end{eqnarray}
The vector FFs associated with isovector quark currents for different physical states satisfy the following relations:
\begin{eqnarray}\label{eq.vff1relt}
&F^{\frac{\bar{u}u - \bar{d}d}{\sqrt2}}_{+,\pi^+\pi^-}=F^{\bar{u}d}_{+,\pi^-\pi^0}=-F^{\bar{d}u}_{+,\pi^+\pi^0}\,,\nonumber \\
&F^{\frac{\bar{u}u - \bar{d}d}{\sqrt2}}_{+,K^0\bar{K}^0}=-F^{\frac{\bar{u}u - \bar{d}d}{\sqrt2}}_{+,K^+K^-}=-\frac{1}{\sqrt{2}}F^{\bar{u}d}_{+,K^0K^-}=-\frac{1}{\sqrt{2}}F^{\bar{d}u}_{+,K^+\bar{K}^0}\,. 
\end{eqnarray}
And for the vector quark currents with $I=1/2$, we have the following relationships 
\begin{eqnarray}\label{eq.vff1d2relt}
&F^{\bar{s}u}_{+,K^+\pi^0}=\frac{1}{\sqrt{2}}F^{\bar{s}u}_{+,K^0\pi^+}=-F^{\bar{u}s}_{+,K^-\pi^0}=-\frac{1}{\sqrt{2}}F^{\bar{u}s}_{+,\bar{K}^0\pi^-}\nonumber\\
&=-F^{\bar{s}d}_{+,K^0\pi^0}=\frac{1}{\sqrt{2}}F^{\bar{s}d}_{+, K^+\pi^-}=F^{\bar{d}s}_{+,\bar{K}^0\pi^0}=-\frac{1}{\sqrt{2}}F^{\bar{d}s}_{+,K^-\pi^+},\nonumber\\
&F^{\bar{s}u}_{+,K^+\eta^{(')}}=-F^{\bar{u}s}_{+,K^-\eta^{(')}}=F^{\bar{s}d}_{+,K^0\eta^{(')}}=-F^{\bar{d}s}_{+,\bar{K}^0\eta^{(')}}\,.
\end{eqnarray}

\subsection{Calculation of the tensor form factors}

The tensor FFs are defined by the following matrix elements of the tensor currents 
\begin{equation}
\langle 0 | \bar{q}^m \sigma^{\mu\nu} q^n |P(p_1)Q(p_2) \rangle \equiv i\frac{\Lambda_2^{T}}{F_\pi^2}(p_1^\mu p_2^\nu-p_1^\nu p_2^\mu) F^{mn}_{T,PQ}(s)\,,
\end{equation}
where $F^{mn}_{T,PQ}(s)$ stands for the tensor FFs. After calculating the Feynman diagrams in Fig.~\ref{fig.feyndiag} by taking the external sources as the tensor ones, one can obtain the perturbative results of the various tensor FFs. The relevant two-meson channels coupled to the various tensor quark currents are the same as the vector cases in Table~\ref{tab.vff}. 
The LO results for the seven independent tensor FFs are 
\begin{eqnarray}
&F^{\frac{\bar{u}u+\bar{d}d}{\sqrt2},\text{LO}}_{T,K^+K^-}=-\frac{1}{\sqrt{2}},\quad F^{\bar{s}s,\text{LO}}_{T,K^+K^-}=1 \,,\quad   F^{\bar{u}d,\text{LO}}_{T,\pi^-\pi^0}=-\sqrt{2} \,,\quad F^{\bar{u}d,\text{LO}}_{T,K^0K^-}=-1  \,,\nonumber\\ 
& F^{\bar{u}s,\text{LO}}_{T,K^-\pi^0}=-\frac{1}{\sqrt{2}} \,,\quad F^{\bar{u}s,\text{LO}}_{T,K^-\eta}=-\frac{\sqrt{6}}{2}c_{\theta} \,,\quad F^{\bar{u}s,\text{LO}}_{T,K^-\eta'}=-\frac{\sqrt{6}}{2}s_{\theta} \,.
\end{eqnarray}
The full expressions of these seven independent perturbative tensor FFs can be found in Eqs.~\eqref{TFFbegin}--\eqref{TFFend} in the Appendix. 

To account for the non-perturbative meson-meson interactions, one can use a similar formalism as the vector case in Eq.~\eqref{eq.unifv} to perform the unitarization for the tensor FFs, i.e.,
\begin{equation}\label{eq.unift}
F^I_T(s) = \tilde{Q}(s)^{-1} \cdot \left[ 1 + N^{IJ}(s) \cdot g^{IJ}(s) \right]^{-1} \cdot \tilde{Q}(s) \cdot R_T^I(s)\,,
\end{equation}
with 
\begin{equation}
R_T^I(s) = F_T^I(s)^{(2)+\text{Res+Loop}} + \tilde{Q}(s)^{-1} \cdot N^{IJ}(s)^{(2)}\cdot \, g^{IJ}(s)\cdot \,\tilde{Q}(s) \cdot F_T^I(s)^{(2)}.
\end{equation}
The unitarized tensor FFs in Eq.~\eqref{eq.unift} satisfy the unitarity relations 
\begin{equation}
\text{Im}\, F^I_T(s) = \tilde{Q}(s)^{-1} \cdot T^{IJ}(s)^* \cdot \rho(s) \cdot\tilde{Q}(s) \cdot F^{I}_T(s)\,.
\end{equation}
It is noted that $J=1$ has been taken when building the unitarized tensor FFs.  
The tensor FFs in the isospin bases used in the unitarization procedure are constructed according to the conventions of Eqs.~\eqref{eq.ffisospin0norm}, \eqref{eq.ffisospin1norm}, and \eqref{eq.ffisospin1d2norm}. 
It turns out that the relations among the tensor FFs for the physical states are the same as those in the vector case, given in Eqs.~\eqref{eq.vff0relt}--\eqref{eq.vff1d2relt}; see, for example, the discussion in Ref.~\cite{Hoferichter:2018zwu}.

\section{Phenomenological discussions}\label{sec.pheno}

In this section, we present the phenomenological results for the scalar, vector, and tensor FFs. All the unknown parameters entering the various FFs have already been determined in previous studies of $U(3)$ meson--meson scattering~\cite{Guo:2011pa,Guo:2012yt}, except for $F_V$ in Eq.~\eqref{eq.lagvector}, which enters the vector FFs, and the tensor-current couplings $\Lambda_2^T$ in Eq.~\eqref{eq.lagtc1} and $F_V^T$ in Eq.~\eqref{eq.lagtv}.
For the vector sector, we take $F_V=130\,\mathrm{MeV}$, which is close to the large-$N_C$ estimate $\sqrt{2}F_\pi$~\cite{Ecker:1988te}. For the tensor sector, the couplings are fixed to $\Lambda_2^T=12\,\mathrm{MeV}$ and $F_V^T=F_V/\sqrt{2}$ in order to facilitate a direct comparison with the results of Ref.~\cite{Miranda:2018cpf}.
For all remaining parameters, we adopt the results obtained from the best fit in Ref.~\cite{Guo:2012yt}, including both the central values and the associated uncertainties, as our baseline setup. This choice will be referred to as Fit~I in the following discussions. For comparison purposes, we also consider the parameter set determined in Ref.~\cite{Guo:2011pa}, which we denote as Fit~II.

To be specific, we explicitly give the values of the different parameters in Fit I below
\begin{eqnarray}
&c_d = \left(19.8^{+2.0}_{-5.2}\right) \, \text{MeV}, \quad 
c_m = \left(41.9^{+3.9}_{-9.2}\right) \, \text{MeV}, \quad 
M_{S_8} = \left(1397^{+73}_{-61}\right) \, \text{MeV},\nonumber \\ 
&M_{S_1} = \left(1100^{+30}_{-63}\right) \, \text{MeV}, \quad
M_\rho = \left(801.2^{+8.2}_{-6.9}\right) \, \text{MeV}, \quad 
M_{K^*} = \left(910.0^{+7.0}_{-9.1}\right) \, \text{MeV},\nonumber \\ 
&G_V = \left(62.1^{+1.9}_{-2.1}\right) \, \text{MeV}, \quad
M_0 = \left(951^{+50}_{-50}\right) \, \text{MeV},\quad 
a_{SL}^{00,\pi\pi} = -1.27^{+0.12}_{-0.12}, \nonumber \\ 
&a_{SL}^{\frac{1}{2}0,K\pi} = -1.12^{+0.12}_{-0.17}, \quad 
a_{SL}^{\frac{1}{2}0,K\eta'} = -1.25^{+1.11}_{-1.23},  \quad
a_{SL}^{10,\pi\eta} = 2.0^{+3.3}_{-4.5},  \quad
a_{SL}^{00,K\bar{K}} = -0.95^{+0.33}_{-0.16}, \nonumber \\ 
&a_{SL}^{\frac{1}{2}0,K\eta} = -0.08^{+0.38}_{-1.04}, \quad
\delta L_8 = 0.23^{+0.29}_{-0.19} \times 10^{-3},  \quad 
\Lambda_2 = -0.37^{+0.19}_{-0.19}\,, \label{eq:group1}
\end{eqnarray}
and see Ref.~\cite{Guo:2012yt} for details about the way they are determined. It should be noted that we use $M_\rho$ to estimate the numerical value of the parameter $M_V$.
The large $N_C$ relation is imposed for the scalar resonance couplings in this fit, namely $\tilde{c}_d=\frac{c_d}{\sqrt{3}}$ and $\tilde{c}_m=\frac{c_m}{\sqrt{3}}$. The parameters related to the pseudoscalar resonances are fixed at $d_m=\sqrt{3}\tilde{d}_m=30\,\text{MeV}$ and $M_{P_8}= M_{P_1}=1350\,\text{MeV}$~\cite{Guo:2012yt}.  The relationships between the remaining subtraction constants used in the former reference read
\begin{eqnarray}
a_{SL}^{00, \eta \eta} &=& a_{SL}^{00, \eta \eta'} = a_{SL}^{00, \eta' \eta'} = a_{SL}^{00, K\bar{K}}, \nonumber \\
a_{SL}^{20, \pi \pi} &=& a_{SL}^{00, \pi \pi}, \quad 
a_{SL}^{\frac{3}{2}0, K\pi} = a_{SL}^{\frac{1}{2}0, K\pi}, \nonumber \\
a_{SL}^{10, \pi \eta'} &=& a_{SL}^{10, K\bar{K}} = a_{SL}^{00, K\bar{K}}. \label{eq:relations}
\end{eqnarray}
Moreover, all subtraction constants for the vector channels are taken to be equal to $a_{SL}^{00,\pi\pi}$. Ref.~\cite{Guo:2012yt} also includes an additional multiplet of bare scalar resonances above 2~GeV. Although these states primarily serve as a smooth background contribution in the energy region below 1.5 GeV, they are included in our analysis to maintain consistency with the original fitting setup; we refer the reader to Ref.~\cite{Guo:2012yt} for further details.
For Fit~II, we do not explicitly list the corresponding parameter values here and instead refer to Ref.~\cite{Guo:2011pa} for the complete setup. 

The determinations of the subtraction constants in the isovector scalar scatterings with $\pi\eta$, $K\bar{K}$ and $\pi\eta'$ coupled channels, i.e., $a_{SL}^{10,\pi\eta}=0.56$ and $a_{SL}^{10,K\bar{K}}=a_{SL}^{10,\pi\eta'}=-1.62$, are later updated in Ref.~\cite{Guo:2016zep} by introducing in the fit the precise finite-volume energy levels from the lattice QCD simulations~\cite{Dudek:2016cru}. We will denote such a fit as Fit III in later discussions. 

With the aforementioned parameters from the fits of Refs.~\cite{Guo:2012yt,Guo:2011pa,Guo:2016zep}, a plethora of resonances have been found in the meson-meson scattering amplitudes, including the isoscalar scalars $f_0(500)$, $f_0(980)$ and $f_0(1370)$, the isospin-1/2 scalars $K^*_0(700)$ and $K^*_0(1430)$, the isovector scalars $a_0(980)$ and $a_0(1450)$, and the vectors $\rho(770)$, $K^*(892)$ and $\phi(1020)$. 
Interestingly, these resonances manifest themselves in rather different ways for the various two-meson FFs, as we will discuss below. 

\subsection{Results of the scalar form factors}

\begin{figure}[htbp]
    \centering
    \begin{minipage}[b]{0.41\linewidth}
        \centering
        \includegraphics[width=\linewidth]{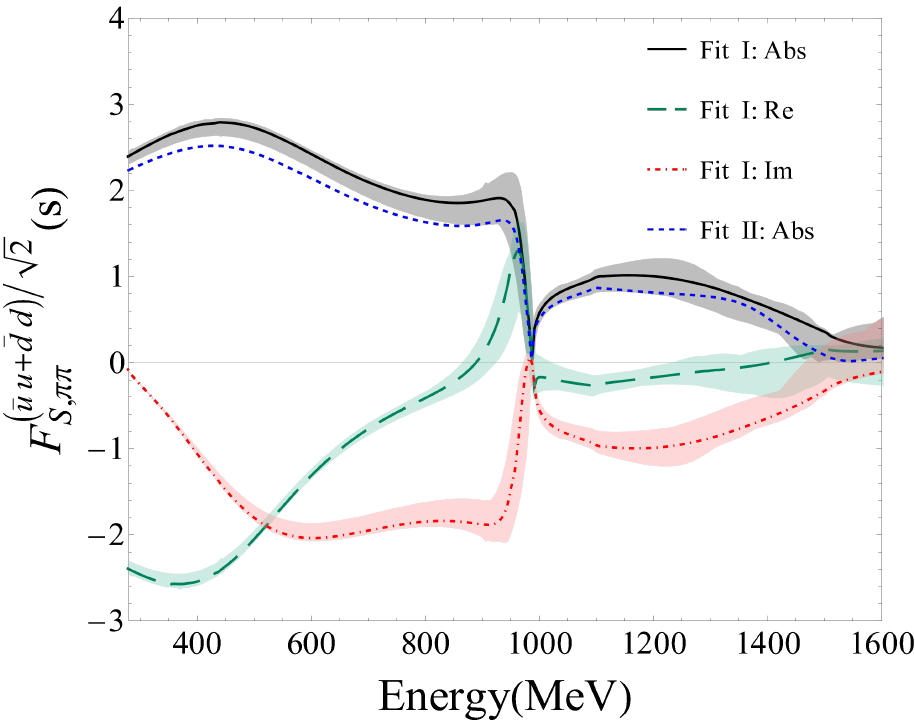}
    \end{minipage}    \hspace{0.05\linewidth}
    \begin{minipage}[b]{0.40\linewidth}
        \centering
        \includegraphics[width=\linewidth]{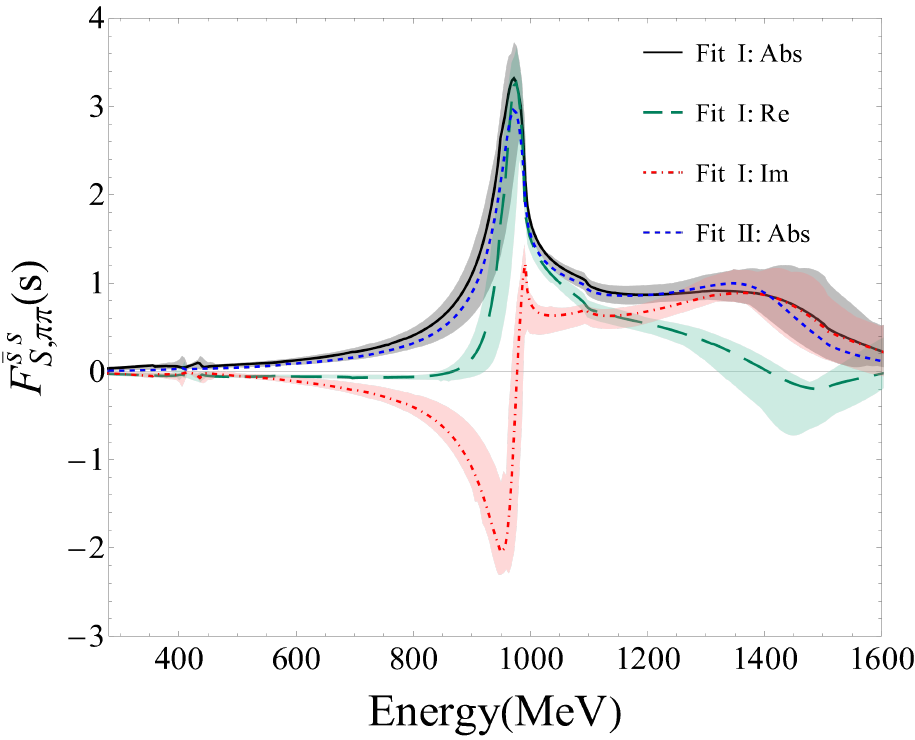}
    \end{minipage}
    \caption{Scalar $\pi\pi$ FFs of the isoscalar quark currents: $F^{(\bar{u}u+\bar{d}d)/\sqrt{2}}_{S,\pi\pi}$ (left) and $F^{\bar{s}s}_{S,\pi\pi}$ (right). 
    In each panel, the black solid curve shows the modulus as a function of the $\pi\pi$ energy, obtained with the parameter set of Fit \uppercase\expandafter{\romannumeral1}. The light-gray shaded areas indicate the corresponding error bands obtained by propagating the parameter uncertainties of Fit~\uppercase\expandafter{\romannumeral1}. The real and imaginary parts from Fit \uppercase\expandafter{\romannumeral1} are denoted by the green long-dashed and red dot-dashed curves, respectively; the light-green and light-pink shaded areas indicate the corresponding error bands, also propagated from the Fit \uppercase\expandafter{\romannumeral1} parameter uncertainties. The blue dashed curve in each panel shows the modulus obtained with the parameter set of Fit \uppercase\expandafter{\romannumeral2}. In order to avoid the overloading of the figures, only the central curves for the modulus from Fit II are provided.  }
    \label{fig.SFFpipichannel}
\end{figure}

For the scalar FFs with the isoscalar quark currents, although the curves have been explicitly provided for the $\pi\pi$ channel in Ref.~\cite{Guo:2012yt}, we repeat the results here for the sake of completeness, and the error bands are further included in Fig.~\ref{fig.SFFpipichannel}. The $f_0(500)$ shows up as the mild bump in $F^{(\bar{u}u+\bar{d}d)/\sqrt{2}}_{S,\pi\pi}$, i.e., the $\pi\pi$ scalar FF of the $(u\bar{u}+d\bar{d})/\sqrt{2}$ current, while it is barely seen in $F^{\bar{s}s}_{S,\pi\pi}$, viz. the $\pi\pi$ scalar FF of the $s\bar{s}$ current. The $f_0(980)$ shows rather different signals in $F^{(\bar{u}u+\bar{d}d)/\sqrt{2}}_{S,\pi\pi}$ and $F^{\bar{s}s}_{S,\pi\pi}$, where a prominent kink structure emerges in the former case and an obvious narrow peak manifests in the latter. Broad bumps appearing in both of them around 1.4~GeV are caused by the $f_0(1370)$ resonance.

\begin{figure}[htbp]
    \centering
    \begin{minipage}[b]{0.405\linewidth}
        \centering
        \includegraphics[width=\linewidth]{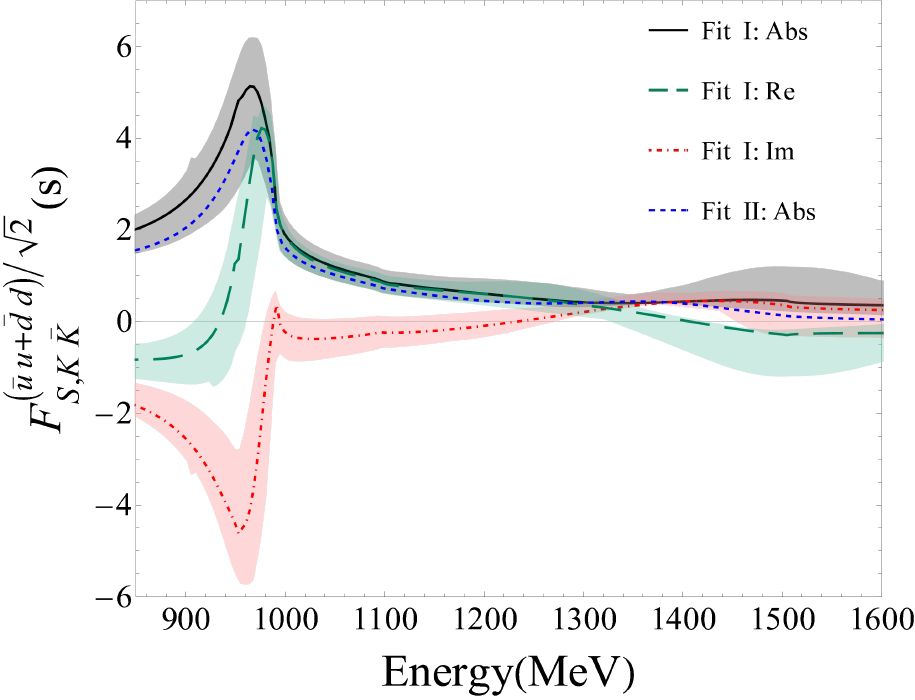}
    \end{minipage}    \hspace{0.05\linewidth}
    \begin{minipage}[b]{0.40\linewidth}
        \centering
        \includegraphics[width=\linewidth]{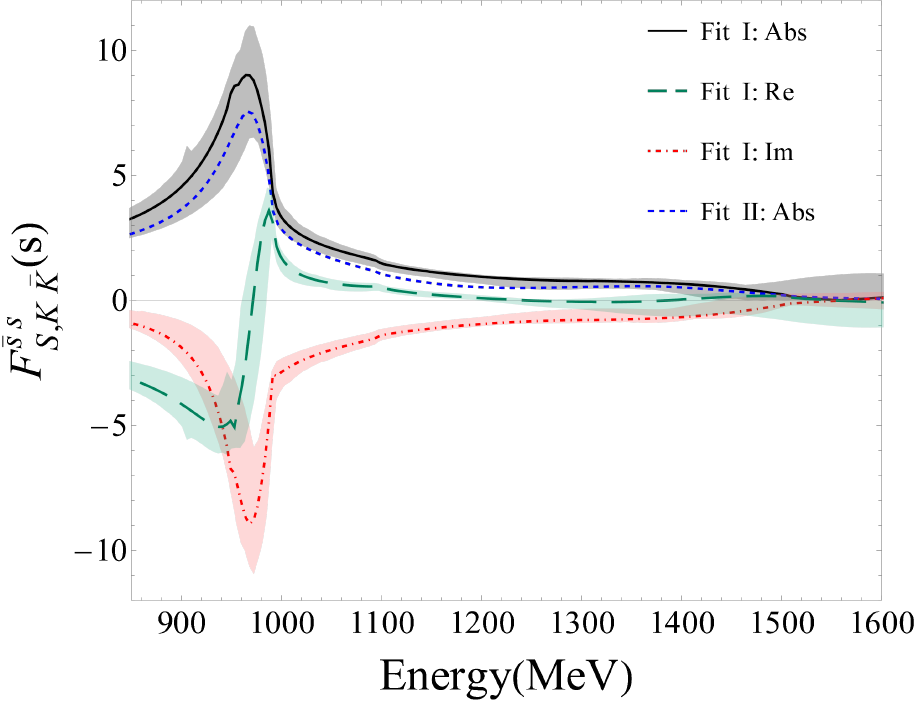}
    \end{minipage}
    \caption{Scalar $K\bar{K}$ FFs of the isoscalar quark currents: $F^{(\bar{u}u+\bar{d}d)/\sqrt{2}}_{S,K\bar{K}}$ (left) and $F^{\bar{s}s}_{S,K\bar{K}}$ (right). The line styles and shaded bands follow the same conventions as in Fig.~\ref{fig.SFFpipichannel}.}
    \label{fig.SFFKKchannel}
\end{figure}

\begin{figure}[htbp]
    \centering
    \begin{minipage}[b]{0.405\linewidth}
        \centering
        \includegraphics[width=\linewidth]{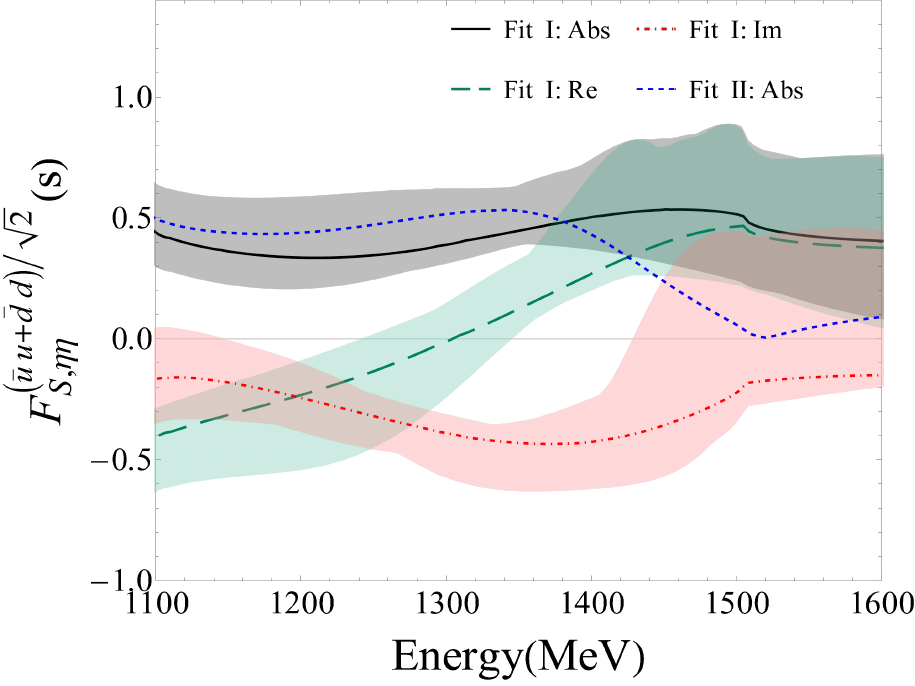}
    \end{minipage}    \hspace{0.05\linewidth}
    \begin{minipage}[b]{0.40\linewidth}
        \centering
        \includegraphics[width=\linewidth]{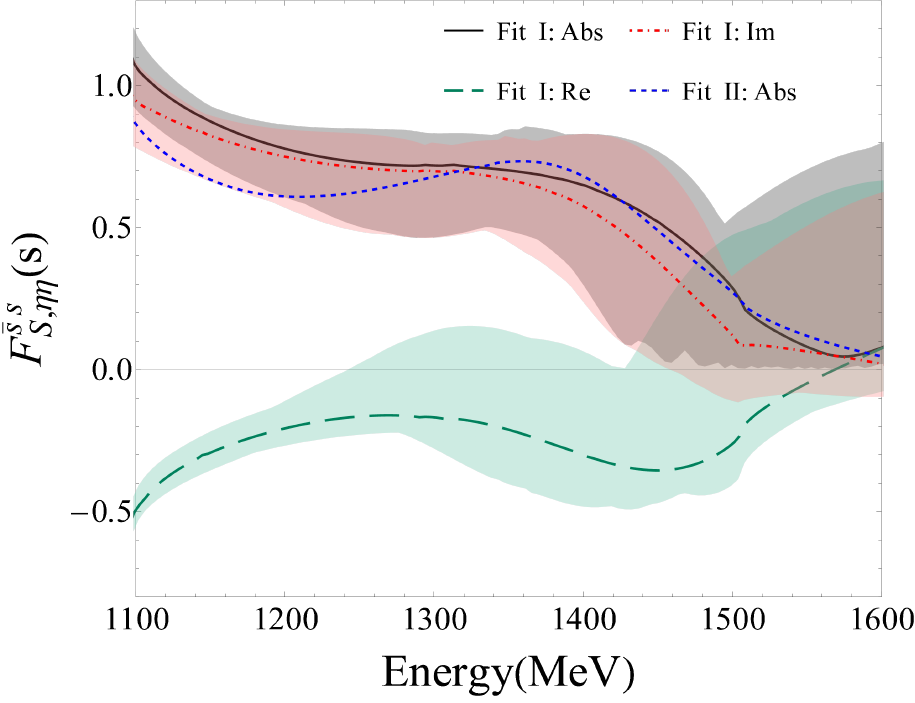}
    \end{minipage}
    \caption{Scalar $\eta\eta$ FFs: $F^{(\bar{u}u+\bar{d}d)/\sqrt{2}}_{S,\eta\eta}$ (left) and $F^{\bar{s}s}_{S,\eta\eta}$ (right). The line styles and shaded bands follow the same conventions as in Fig.~\ref{fig.SFFpipichannel}.}
    \label{fig.SFFetaetachannel}
\end{figure}

The curves corresponding to the $\kkbar$ scalar FFs of the isoscalar quark currents $(u\bar{u}+d\bar{d})/\sqrt{2}$ and $s\bar{s}$ are given in Fig.~\ref{fig.SFFKKchannel}. Prominent peaks around the $\kkbar$ thresholds, where the $f_0(980)$ resonance resides, show up in both FFs. While the $f_0(1370)$ signal is not profound in these two FFs. For the $\eta\eta$ scalar FFs, as illustrated in Fig.~\ref{fig.SFFetaetachannel}, similar magnitudes are seen for $F^{(\bar{u}u+\bar{d}d)/\sqrt{2}}_{S,\eta\eta}$ and $F^{\bar{s}s}_{S,\eta\eta}$.

\begin{figure}[htbp]
  \centering
  \begin{minipage}[b]{0.32\linewidth}
    \centering
    \includegraphics[width=\linewidth]{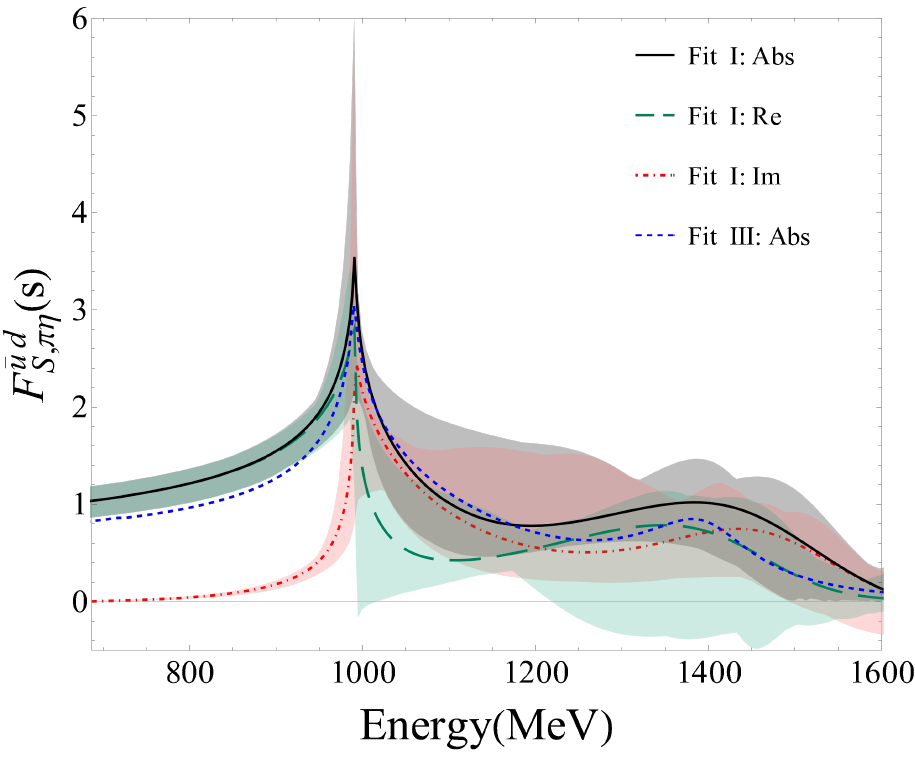}
  \end{minipage}\hfill%
  \begin{minipage}[b]{0.335\linewidth}
    \centering
    \includegraphics[width=\linewidth]{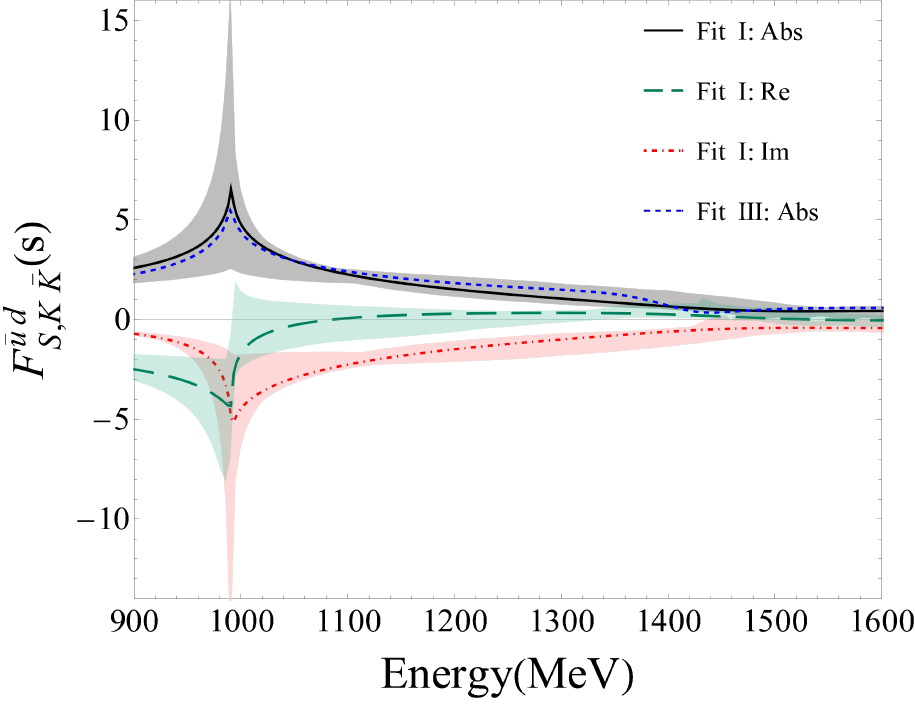}
  \end{minipage}\hfill%
  \begin{minipage}[b]{0.335\linewidth}
    \centering
    \includegraphics[width=\linewidth]{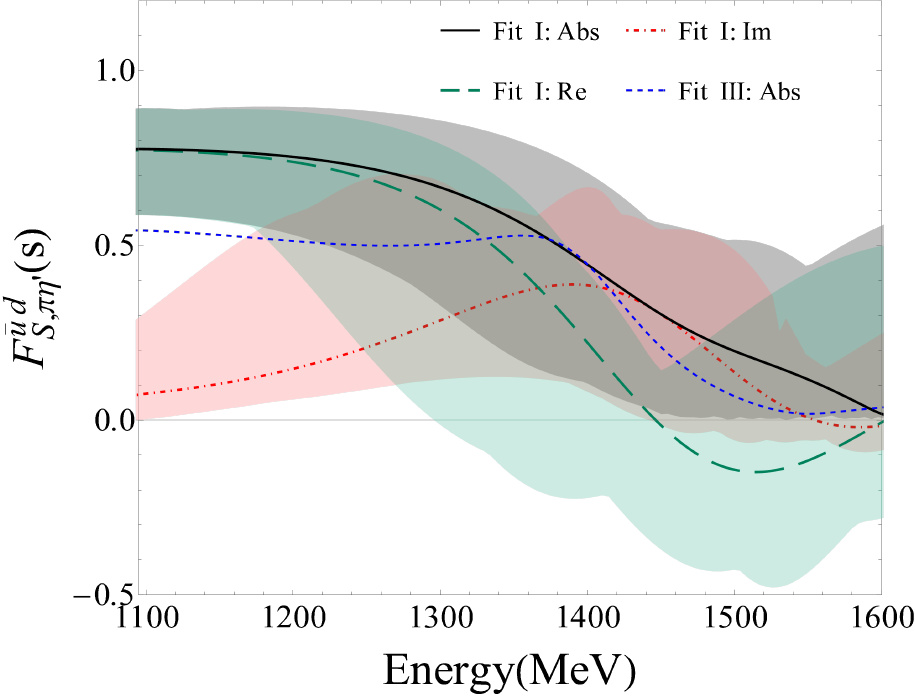}
  \end{minipage}
  \caption{Scalar $\pi\eta$ (left), $\kkbar$ (middle) and $\pi\eta'$ (right) FFs of the isovector quark current $\bar{u}d$: $F^{\bar{u}d}_{S,\pi\eta}$, $F^{\bar{u}d}_{S,K\bar{K}}$, and $F^{\bar{u}d}_{S,\pi\eta'}$. The blue dashed curves show the modulus using the parameter set of the NLO fit obtained in Ref.~\cite{Guo:2016zep}. All other line styles and shaded bands follow the same conventions as in Fig.~\ref{fig.SFFpipichannel}. }
  \label{fig.SFFudchannel}
\end{figure}

The scalar $\pi\eta$, $\kkbar$ and $\pi\eta'$ FFs of the isovector quark current $\bar{u}d$ are illustrated in  Fig.~\ref{fig.SFFudchannel}, where we show the results of Fit I and Fit III, corresponding to the best fits of Refs.~\cite{Guo:2012yt} and \cite{Guo:2016zep}, respectively. For the isovector scalar $\pi\eta$ and $\kkbar$ FFs, the most obvious structures appear around the $\kkbar$ threshold, where the $a_0(980)$ resonance resides. The higher $a_0(1450)$ resonance manifests as a clear bump around 1.4~GeV in the $\pi\eta$ channel, while its signal is not very visible in the $\kkbar$ case. The $\pi\eta'$ FF is smooth around the $\pi\eta'$ threshold, since the $a_0(980)$ resonance lies below this threshold. The excited $a_0(1450)$ does not behave as a bump in the $\pi\eta'$ FF.

\begin{figure}[htbp]
  \centering
  \begin{minipage}[b]{0.325\linewidth}
    \centering
    \includegraphics[width=\linewidth]{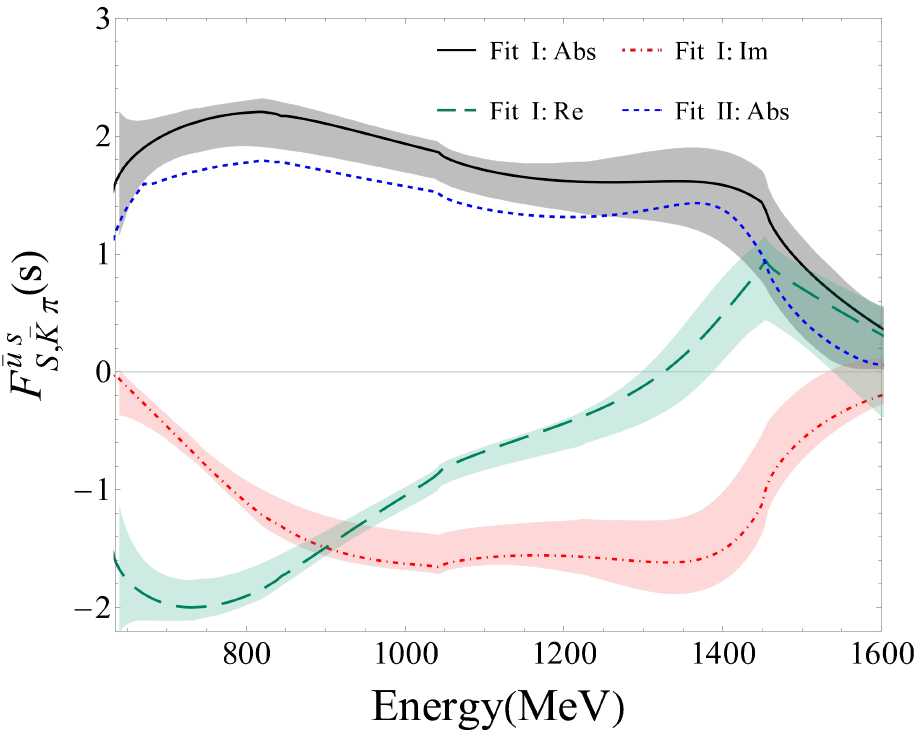}
  \end{minipage}\hfill%
  \begin{minipage}[b]{0.335\linewidth}
    \centering
    \includegraphics[width=\linewidth]{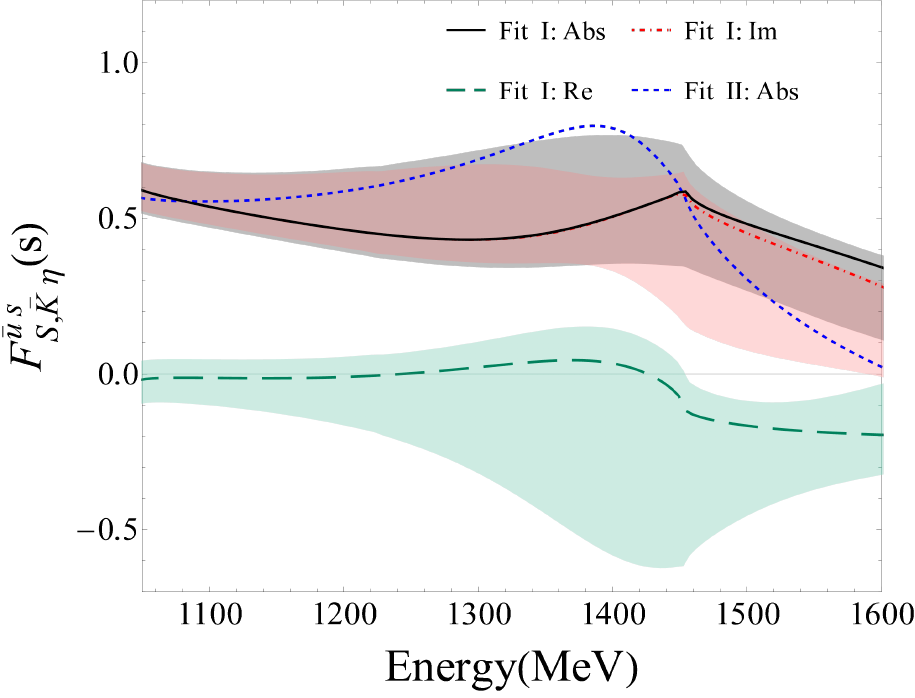}
  \end{minipage}\hfill%
  \begin{minipage}[b]{0.325\linewidth}
    \centering 
    \includegraphics[width=\linewidth]{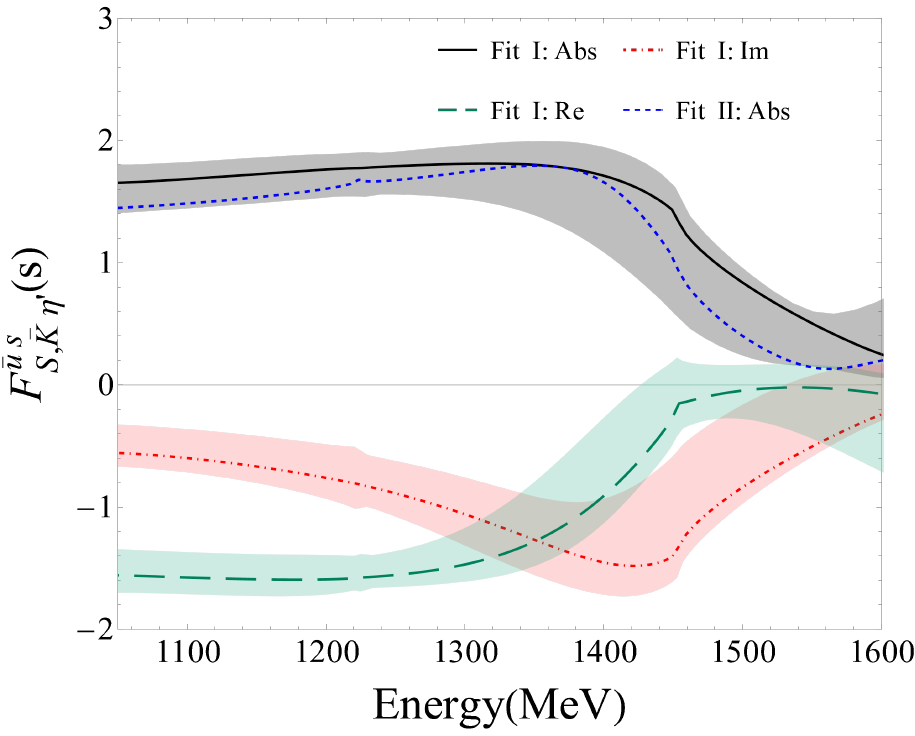}
  \end{minipage}
  \caption{Scalar $\bar{K}\pi$ (left), $\bar{K}\eta$ (middle) and $\bar{K}\eta'$ (right) FFs of the isospin-1/2 quark current $\bar{u}s$: $F^{\bar{u}s}_{S,\bar{K}\pi}$, $F^{\bar{u}s}_{S,\bar{K}\eta}$ and $F^{\bar{u}s}_{S,\bar{K}\eta'}$. The line styles and shaded bands follow the same conventions as in Fig.~\ref{fig.SFFpipichannel}.}
  \label{fig.SFFuschannel}
\end{figure}

The $\bar{K}\pi$, $\bar{K}\eta$ and $\bar{K}\eta'$ scalar FFs of the isospin-1/2 quark current $\bar{u}s$ are exhibited in Fig.~\ref{fig.SFFuschannel}. The magnitude trends of the three different FFs with respect to the two-meson energies seem roughly similar; namely, the excited scalar $K^*_0(1430)$ manifests like a shoulder around 1.4~GeV in all of them. The low-lying broad $K^*_0(700)$ scalar resonance is not very apparent in the $\bar{K}\pi$ FF. 

\subsection{Results of the vector form factors}

\begin{figure}[h]
    \centering
    \begin{minipage}[b]{0.4\linewidth}
        \centering
        \includegraphics[width=\linewidth]{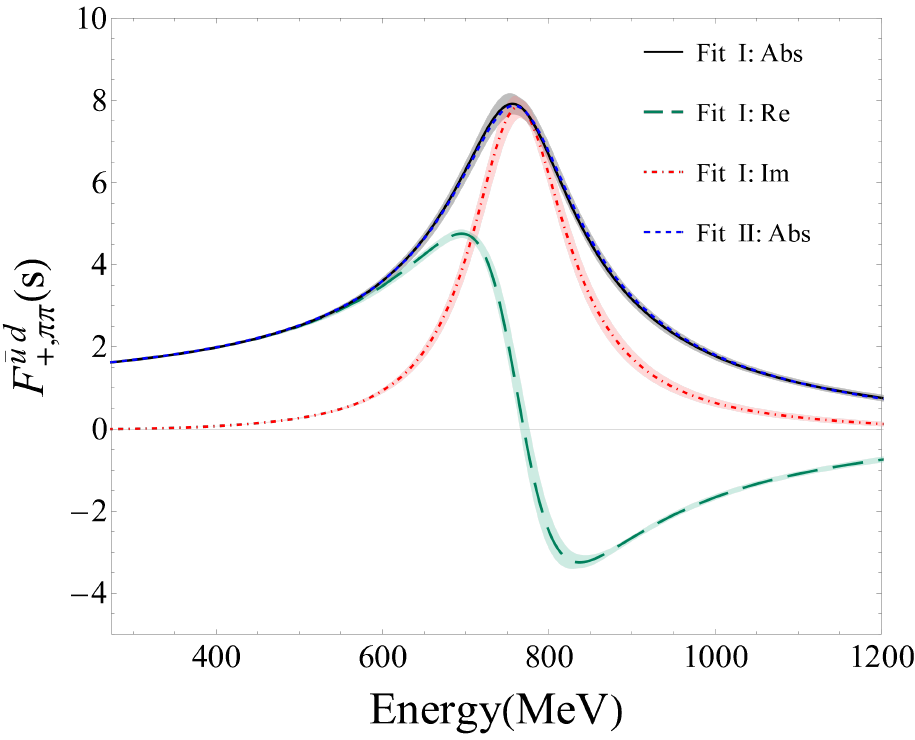}
    \end{minipage}
    \hspace{0.05\linewidth}
    \begin{minipage}[b]{0.4\linewidth}
        \centering
        \includegraphics[width=\linewidth]{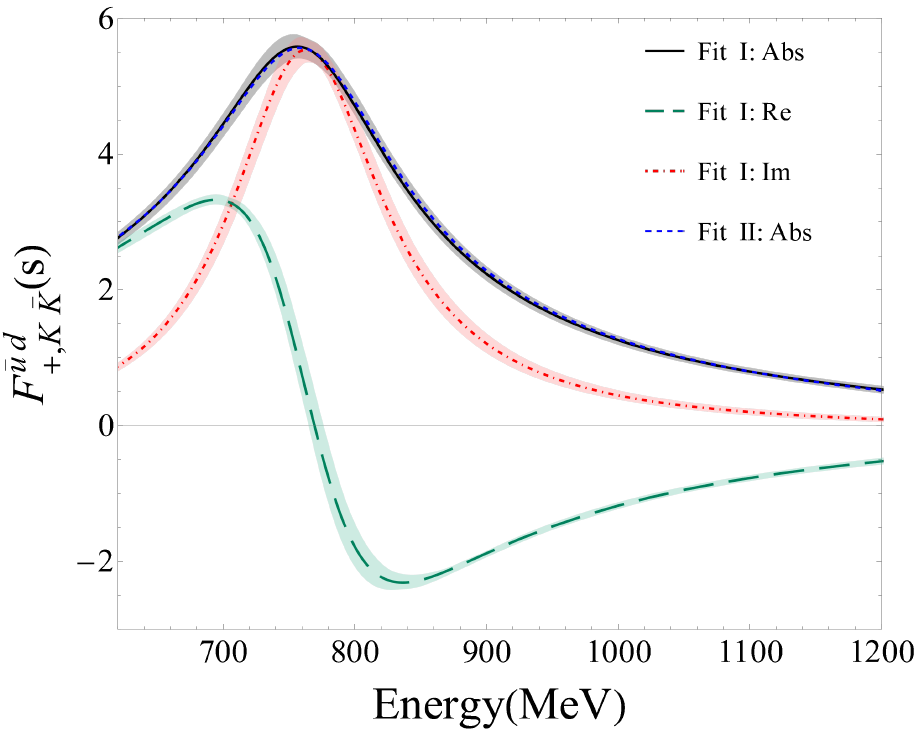}
    \end{minipage}
    \caption{Vector $\pi\pi$ (left) and $\kkbar$ (right) FFs of the isovector quark current $\bar{u}\gamma^\mu d$: $F^{\bar{u}d}_{+,\pi\pi}$ and $F^{\bar{u}d}_{+,K\bar{K}}$. The line styles and shaded bands follow the same conventions as in Fig.~\ref{fig.SFFpipichannel}.}
    \label{fig.VFFudchannel}
\end{figure}

In this section, we perform the phenomenological analyses of the various two-meson vector FFs. The plots corresponding to the $\pi\pi$ and $\kkbar$ vector FFs of the isovector quark current $\bar{u}\gamma^\mu d$ are illustrated in  Fig.~\ref{fig.VFFudchannel}. The $\rho(770)$ resonance peak is prominent and also dominates the line shapes of both $F^{\bar{u}d}_{+,\pi\pi}$ and $F^{\bar{u}d}_{+,K\bar{K}}$ in the focused energy region below 1.2~GeV, though the $\rho(770)$ lies far below the $\kkbar$ threshold. 

\begin{figure}[h]
  \centering
  \begin{minipage}[b]{0.4\linewidth}
    \centering
    \includegraphics[width=\linewidth]{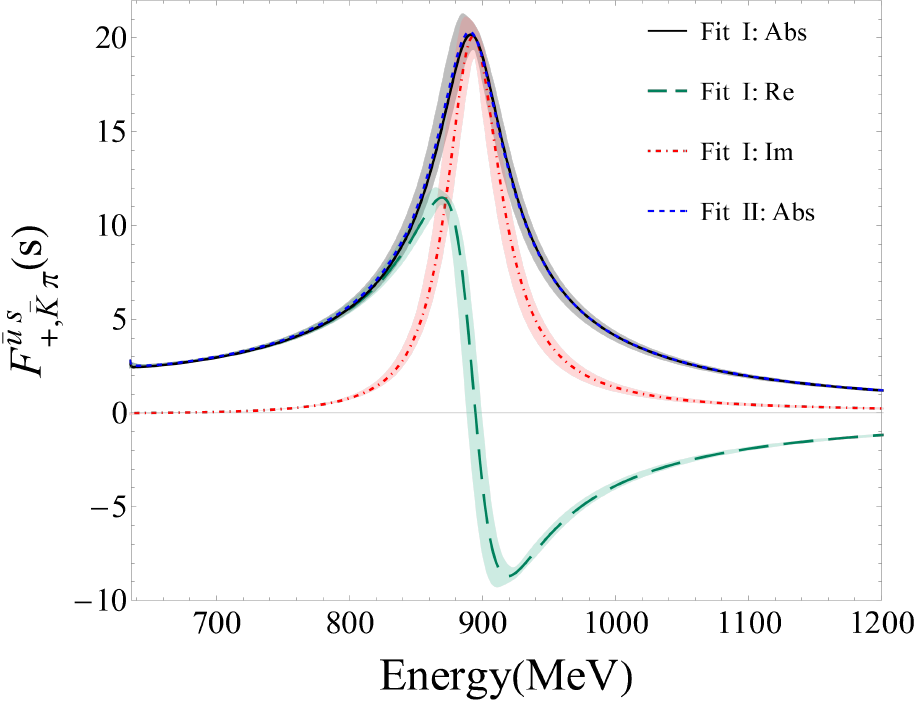}
  \end{minipage}    \hspace{0.05\linewidth}
  \begin{minipage}[b]{0.4\linewidth}
    \centering
    \includegraphics[width=\linewidth]{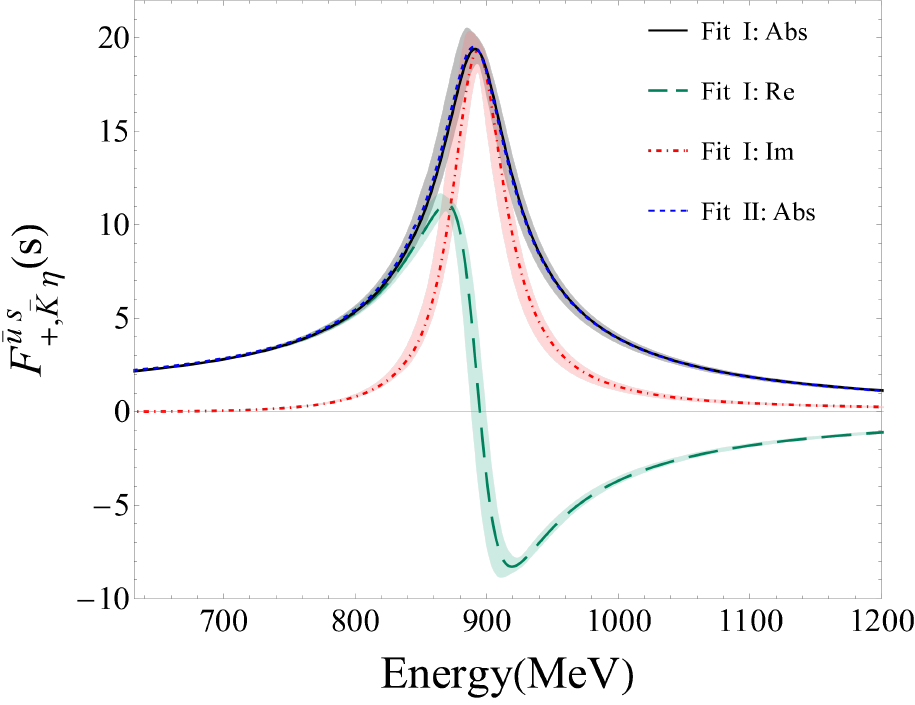}
  \end{minipage}\hfill%
  \caption{Vector $\bar{K}\pi$ (left) and $\bar{K}\eta$ (right) FFs of the isospin-1/2 quark current $\bar{u}\gamma^\mu s$: $F^{\bar{u}s}_{+,\bar{K}\pi}$ and $F^{\bar{u}s}_{+,\bar{K}\eta}$. The line styles and shaded bands follow the same conventions as in Fig.~\ref{fig.SFFpipichannel}.}
  \label{fig.VFFuschannel}
\end{figure}

The curves representing the vector $\bar{K}\pi$ and $\bar{K}\eta$ FFs of the isospin-1/2 quark current $\bar{u}\gamma^\mu s$ are given in Fig.~\ref{fig.VFFuschannel}. Similarly, the $K^*(892)$ resonance peak is noticeable and rules the spectra of both $F^{\bar{u}s}_{+,\bar{K}\pi}$ and $F^{\bar{u}s}_{+,\bar{K}\eta}$ in the focused energy region below 1.2~GeV, albeit the $K^*(892)$ lies below the $\bar{K}\eta$ threshold. 

\begin{figure}[h]
  \centering
  \begin{minipage}[b]{0.4\linewidth}
    \centering
    \includegraphics[width=\linewidth]{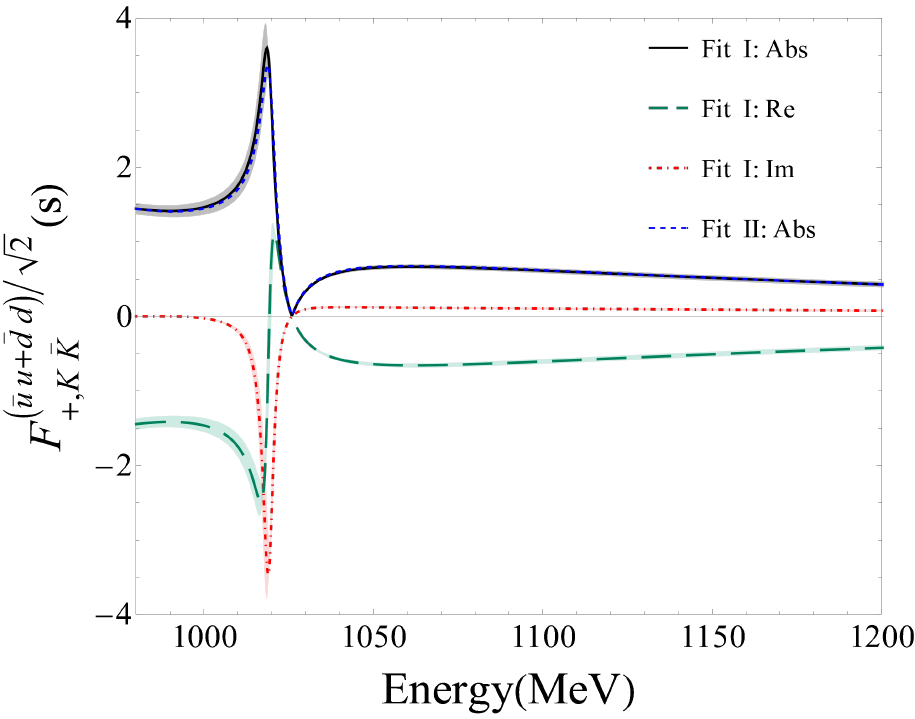}
  \end{minipage}    \hspace{0.05\linewidth}
  \begin{minipage}[b]{0.405\linewidth}
    \centering
    \includegraphics[width=\linewidth]{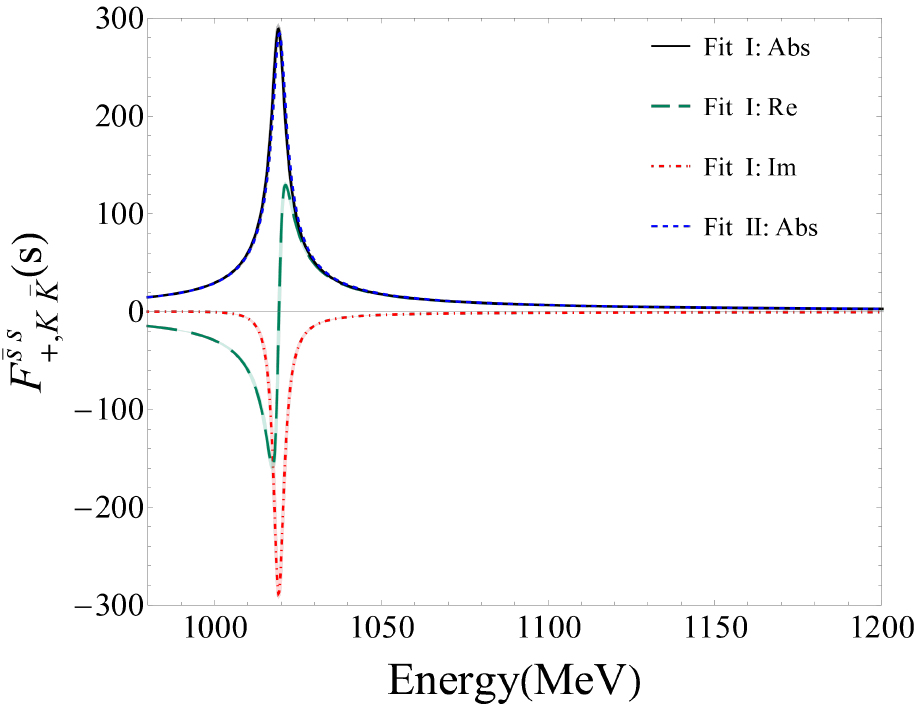}
  \end{minipage}\hfill%
  \caption{Vector $K\bar{K}$ FFs of the isoscalar quark currents $(\bar{u}\gamma^\mu u+\bar{d}\gamma^\mu d)/\sqrt{2}$ (left) and $\bar{s}\gamma^\mu s$ (right): $F^{(\bar{u}u+\bar{d}d)/\sqrt{2}}_{+,K\bar{K}}$ and $F^{\bar{s}s}_{+,K\bar{K}}$. The line styles and shaded bands follow the same conventions as in Fig.~\ref{fig.SFFpipichannel}.}
  \label{fig.VFFuuddsschannel}
\end{figure}

In Fig.~\ref{fig.VFFuuddsschannel}, we present the results of the vector $K\bar{K}$ FFs associated with the isoscalar quark currents $(\bar{u}\gamma^\mu u+\bar{d}\gamma^\mu d)/\sqrt{2}$ and $\bar{s}\gamma^\mu s$. In the case of $F^{\bar{s}s}_{+,K\bar{K}}$, a pronounced $\phi(1020)$ symmetric peak appears and provides the dominant contribution in the energy region below $1.2~\mathrm{GeV}$. By contrast, an asymmetric peak around $1~\mathrm{GeV}$ is observed in $F^{(\bar{u}u+\bar{d}d)/\sqrt{2}}_{+,K\bar{K}}$; also, a zero appears just above $1~\mathrm{GeV}$, a feature not seen in the other vector FFs.

\subsection{Results of the tensor form factors}

\begin{figure}[htbp]
    \centering
    \begin{minipage}[b]{0.4\linewidth}
        \centering
        \includegraphics[width=\linewidth]{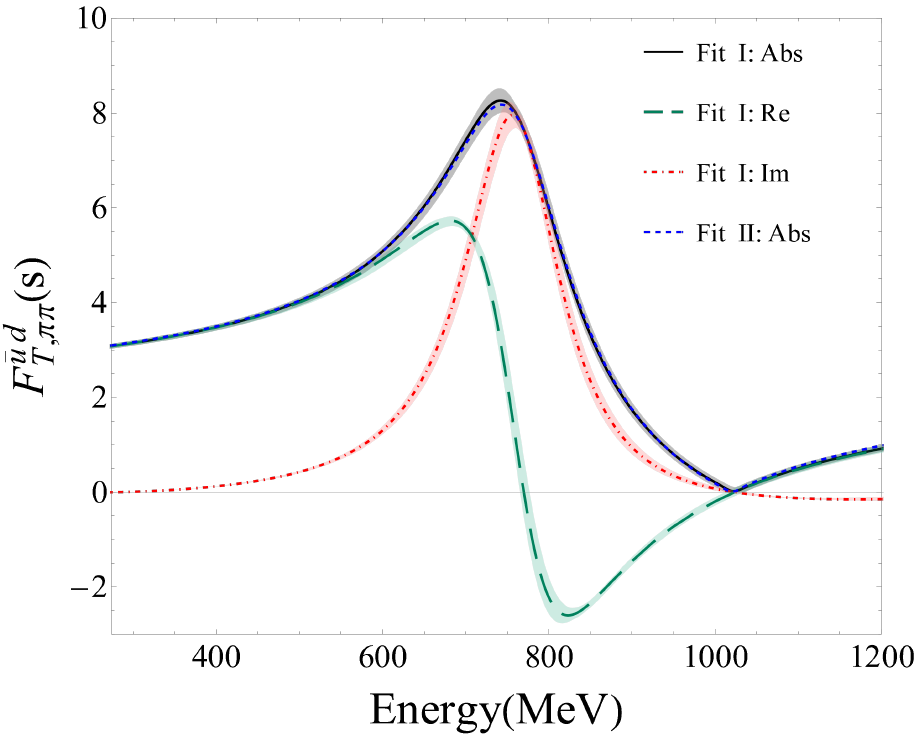}
    \end{minipage} 
    \hspace{0.05\linewidth}
    \begin{minipage}[b]{0.4\linewidth}
        \centering
        \includegraphics[width=\linewidth]{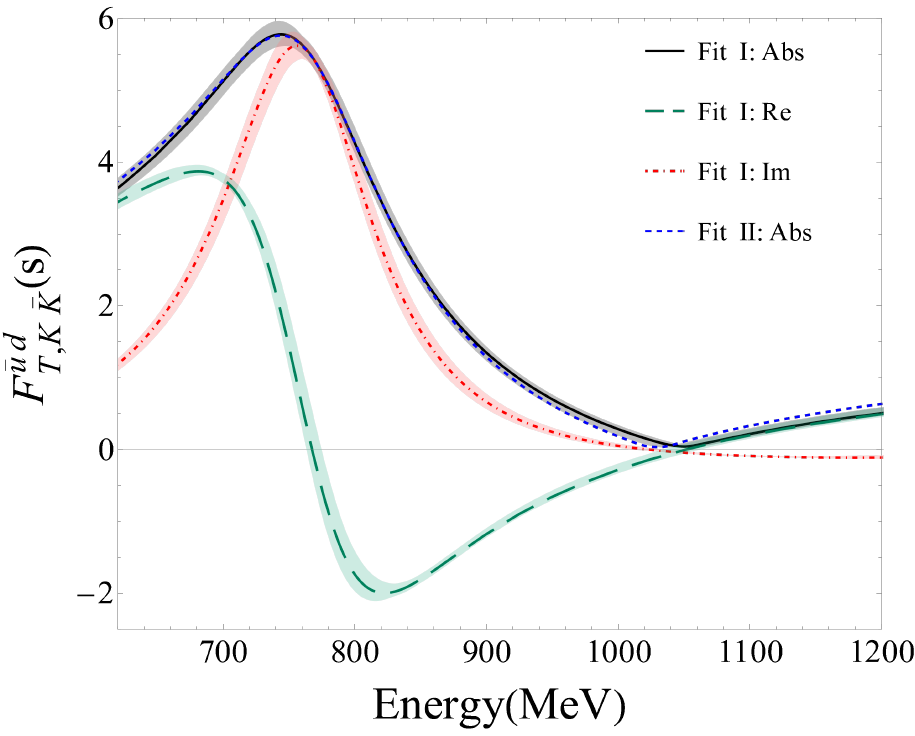}
    \end{minipage}
    \caption{Tensor $\pi\pi$ (left) and $\kkbar$ (right) FFs of the isovector quark current $\bar{u}\sigma^{\mu\nu} d$: $F^{\bar{u}d}_{T,\pi\pi}$ and $F^{\bar{u}d}_{T,K\bar{K}}$. The line styles and shaded bands follow the same conventions as in Fig.~\ref{fig.SFFpipichannel}.  }
    \label{fig.TFFudchannel}
\end{figure}

\begin{figure}[htbp]
    \centering
        \includegraphics[width=0.6\linewidth]{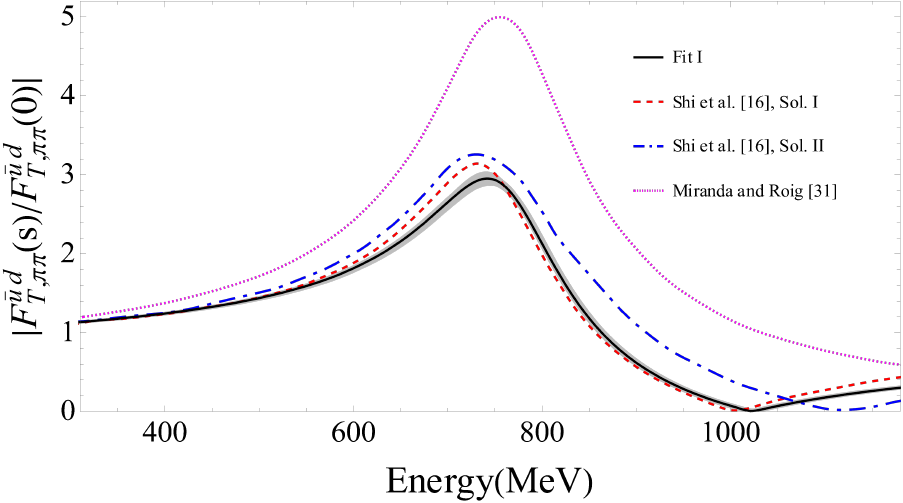}
    \caption{Comparison of the normalized $|F^{\bar{u}d}_{T,\pi\pi}(s)/F^{\bar{u}d}_{T,\pi\pi}(0)|$ obtained using different approaches. The black solid line shows the result of Fit~I from the present work, while the light gray shaded band indicates the uncertainty range obtained from the parameter errors of Fit~I. The red dashed line and the blue dash-dot line represent the curves obtained using the parameter values from Original Fit and Fit~2 in Ref.~\cite{Shi:2020rkz}, respectively, and are labeled as Sol.~I and Sol.~II in the figure. The magenta dotted line corresponds to the result of Ref.~\cite{Miranda:2018cpf}, obtained using dispersive representation.}
    \label{fig.TFFudpipicompared}
\end{figure}

In this subsection, we analyze the phenomenological consequences of the two-meson tensor FFs. In the energy region considered here, below approximately 1.2~GeV, the dominant underlying dynamics governing the various tensor FFs is similar to the vector ones, since they are largely ruled by the ground-state vector resonances. However, for the isovector and isospin-$\frac{1}{2}$ channels, some obvious differences are evident between the vector and tensor FFs, being the presence of the zeros in the latter case above 1~GeV. 

In Fig.~\ref{fig.TFFudchannel}, we show the plots of the $\pi\pi$ and $\kkbar$ tensor FFs for the isovector quark current $\bar{u}\sigma^{\mu\nu} d$, both of which are dominated by the $\rho(770)$ for $E<1.2$~GeV. It is noted that the two FFs exhibit zeros in the energy region around 1~GeV. In Fig.~\ref{fig.TFFudpipicompared}, we compare the normalized quantity $|F^{\bar{u}d}_{T,\pi\pi}(s)/F^{\bar{u}d}_{T,\pi\pi}(0)|$ from our study with those in Refs.~\cite{Miranda:2018cpf,Shi:2020rkz}. The magnitude obtained in Ref.~\cite{Miranda:2018cpf} is clearly larger than others. The results of Ref.~\cite{Shi:2020rkz} from the $SU(3)$ inverse amplitude method (IAM) by taking different parameters look roughly similar to ours from the unitarized $\rxt$ amplitude. However, differences are also seen: (1) the magnitudes of the FF in Ref.~\cite{Shi:2020rkz} are a bit larger than ours; (2) slight variations are observed for the positions of the zeros in $F^{\bar{u}d}_{T,\pi\pi}(s)$. Regarding the quantity of $F^{\bar{u}d}_{T,\pi\pi}(0)$, somewhat different values arise from different approaches. E.g., our prediction of $F^{\bar{u}d}_{T,\pi\pi}(0)=2.9\pm 0.1$ is larger than the value of 1.8 from the IAM calculation in Ref.~\cite{Shi:2020rkz} and the LO value of 1.4 used by the dispersive method in Ref.~\cite{Miranda:2018cpf}. Nevertheless, it is pointed out that the result of $F^{\bar{u}d}_{T,\pi\pi}(0)$ in the IAM approach~\cite{Shi:2020rkz} depends on the poorly determined $O(p^6)$ LECs, which in principle bear rather large uncertainties. 

\begin{figure}[h]
  \centering
  \begin{minipage}[b]{0.4\linewidth}
    \centering
    \includegraphics[width=\linewidth]{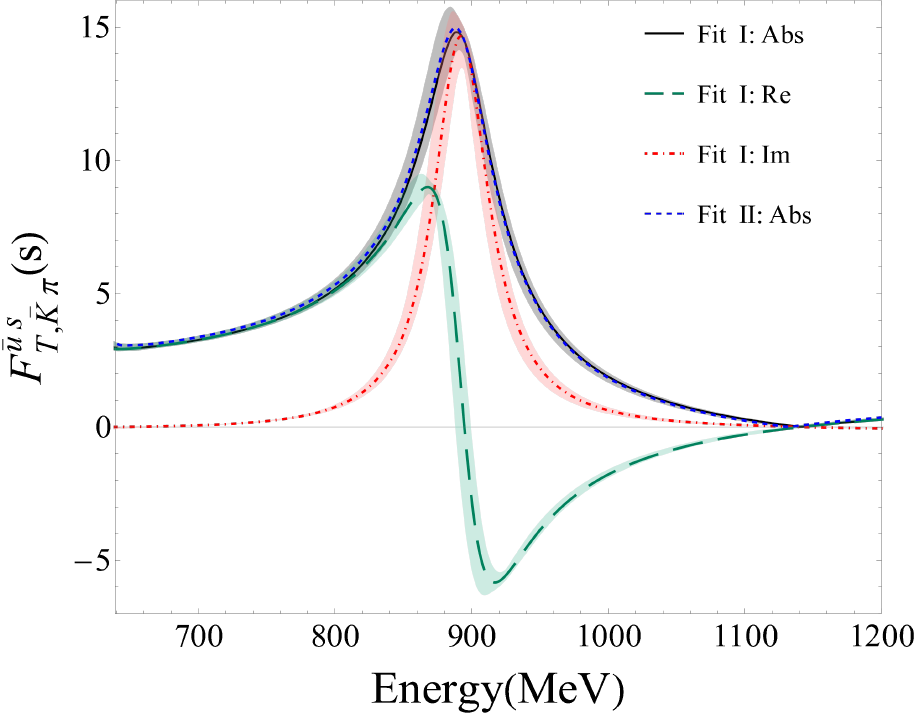}
  \end{minipage}    \hspace{0.05\linewidth}
  \begin{minipage}[b]{0.4\linewidth}
    \centering
    \includegraphics[width=\linewidth]{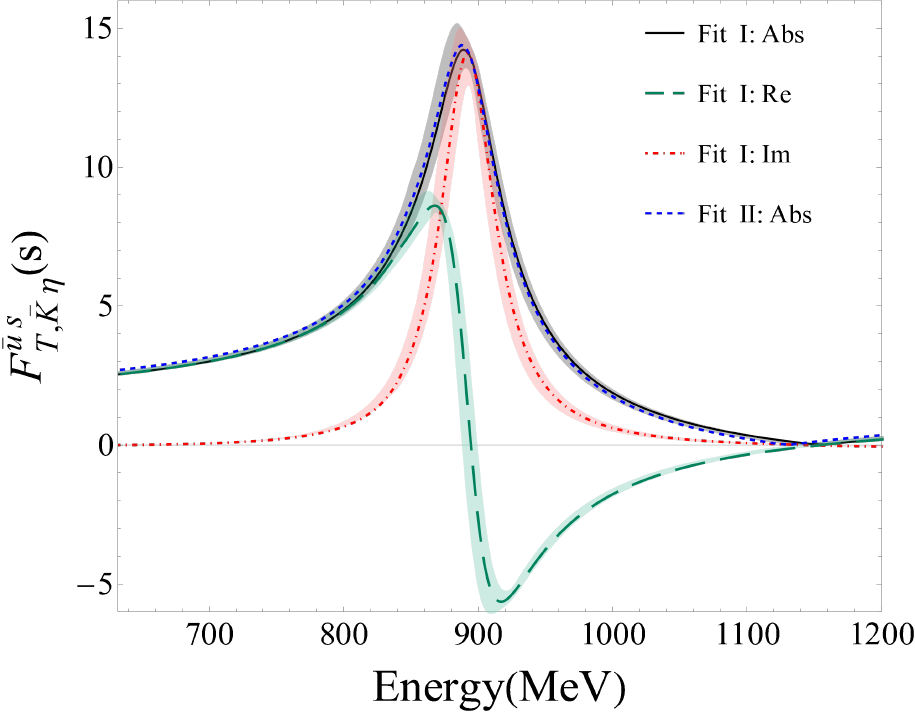}
  \end{minipage}\hfill%
  \caption{Tensor $\bar{K}\pi$ (left) and $\bar{K}\eta$ (right) FFs of the isospin-1/2 quark current $\bar{u}\sigma^{\mu\nu} s$: $F^{\bar{u}s}_{T,\bar{K}\pi}$ and $F^{\bar{u}s}_{T,\bar{K}\eta}$. The line styles and shaded bands follow the same conventions as in Fig.~\ref{fig.SFFpipichannel}.}
  \label{fig.TFFuschannel}
\end{figure}

In Fig.~\ref{fig.TFFuschannel}, we illustrate the results of the $\bar{K}\pi$ and $\bar{K}\eta$ tensor FFs for the isospin-1/2 quark current $\bar{u}\sigma^{\mu\nu} s$, both of which are mostly ruled by the $K^*(892)$ resonance for $E<1.2$~GeV. The positions of the zeros in these two FFs are located around 1.1~GeV. 

\begin{figure}[h]
    \centering
    \begin{minipage}[b]{0.405\linewidth}
        \centering
        \includegraphics[width=\linewidth]{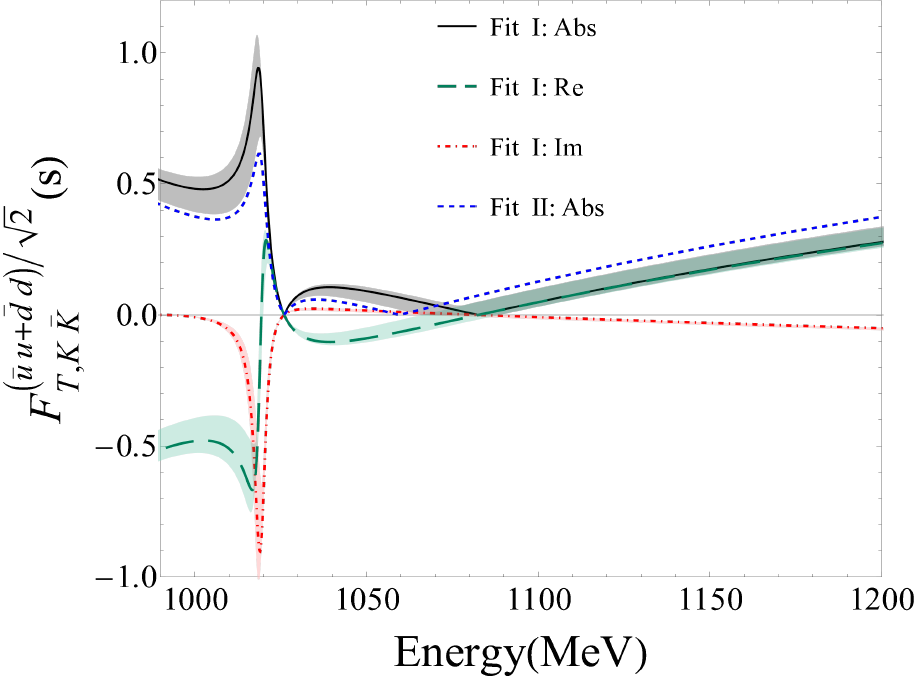}
    \end{minipage}
    \hspace{0.05\linewidth}
    \begin{minipage}[b]{0.4\linewidth}
        \centering
        \includegraphics[width=\linewidth]{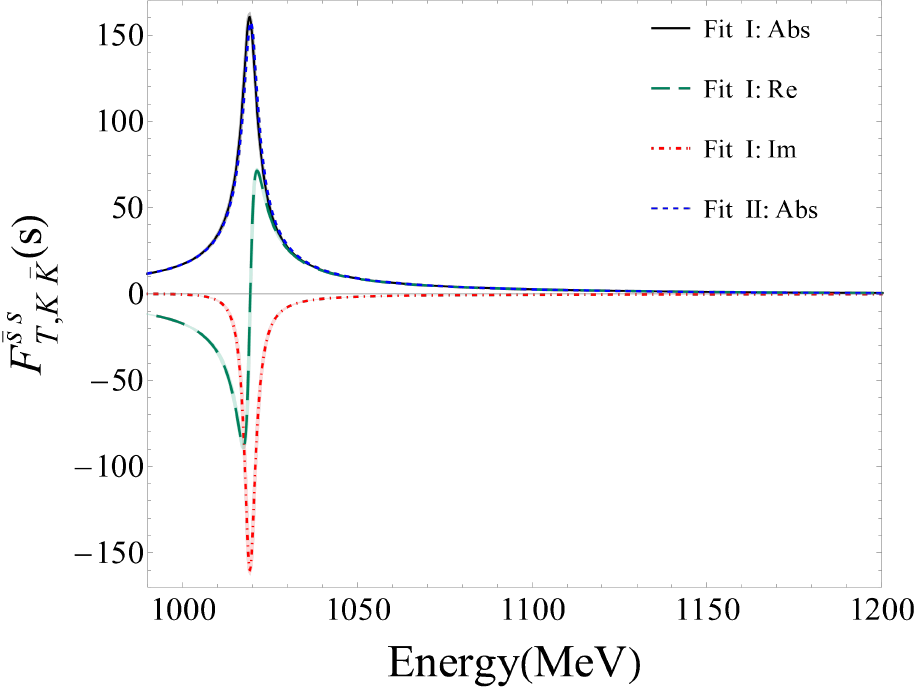}
    \end{minipage}
    \caption{Tensor $K\bar{K}$ FFs of the isoscalar quark currents $(\bar{u}\sigma^{\mu\nu}u+\bar{d}\sigma^{\mu\nu}d)/\sqrt{2}$ (left) and $\bar{s}\sigma^{\mu\nu}s$ (right): $F^{(\bar{u}u+\bar{d}d)/\sqrt{2}}_{T,K\bar{K}}$ and $F^{\bar{s}s}_{T,K\bar{K}}$. The line styles and shaded bands follow the same conventions as in Fig.~\ref{fig.SFFpipichannel}.}
    \label{fig.TFFuuddsschannel}
\end{figure}

Fig.~\ref{fig.TFFuuddsschannel} displays the tensor FFs for the $K\bar{K}$ channel induced by the isoscalar quark currents $(\bar{u}\sigma^{\mu\nu}u+\bar{d}\sigma^{\mu\nu}d)/\sqrt{2}$ and $\bar{s}\sigma^{\mu\nu}s$. A clear symmetric resonance peak from the $\phi(1020)$ is found in $F^{\bar{s}s}_{T,K\bar{K}}$, which dominates the behavior of this FF throughout the region below $1.2~\mathrm{GeV}$. On the other hand, an asymmetric peak structure is seen in $F^{(\bar{u}u+\bar{d}d)/\sqrt{2}}_{T,K\bar{K}}$ around $1~\mathrm{GeV}$. This latter tensor FF also develops a zero in the energy range of $1\sim 1.1$~GeV.

\FloatBarrier

\section{Summary and conclusions}\label{sec.sum}

In this work, we have performed a thorough study of the two-meson FFs of light pseudoscalar mesons within the framework of $U(3)$ resonance chiral theory. At the perturbative level, we have calculated the complete two-meson scalar, vector, and tensor FFs for both the strangeness-conserving and strangeness-changing channels by taking into account both the tree-level resonance-exchange contributions and the pNGB one-loop contributions. 

Next, to account for the meson-meson final-state interactions, we have further constructed the unitarized scalar, vector, and tensor FFs by matching the unitarized meson-meson scattering amplitudes to the perturbative results. An important feature of this approach is that the same dynamical ingredients governing meson-meson scattering also play a crucial role in determining the final-state interactions in the FFs. Therefore, once the scattering amplitudes are constrained by the experimental data and relevant event distributions, most of the parameters entering the FFs can be fixed accordingly, so that many of the FFs are already strongly constrained and can, in many cases, be predicted. 

In the numerical calculations, we have mainly employed parameter sets from previous $U(3)$ meson-meson scattering studies, using the best-fit values from Ref.~\cite{Guo:2012yt} as the baseline input. Based on this fit, all the FFs have been evaluated, and the corresponding uncertainty bands have been provided. For comparison, we have also evaluated the FFs using the parameter values given in Refs.~\cite{Guo:2011pa,Guo:2016zep}. Phenomenologically, we have discussed the main features of various scalar, vector, and tensor FFs in the isoscalar, isovector, and isospin-$1/2$ channels. In the scalar sector, the scalar resonances $f_0(500)$, $f_0(980)$, $f_0(1370)$, $a_0(980)$, $a_0(1450)$, $K_0^*(700)$, and $K_0^*(1430)$ manifest themselves rather differently depending on the quark current and final-state channel. In the vector sector, the low-energy behavior below $1.2$~GeV is largely governed by the ground-state vector resonances, with the $\rho(770)$ dominating the isovector $\pi\pi$ and $K\bar K$ FFs, the $K^*(892)$ controlling the $ K\pi$ and $ K\eta$ channels, and the $\phi(1020)$ producing a pronounced enhancement in the vector FFs of the isoscalar $K \bar K$ channel. The tensor FFs exhibit a resonance pattern qualitatively similar to the vector case, again being mainly controlled by the $\rho(770)$, $K^*(892)$, and $\phi(1020)$ in the low-energy region. We have further compared our $\pi\pi$ tensor FF $F^{\bar{u}d}_{T,\pi\pi}(s)$ with previous results obtained using a dispersive approach and the inverse amplitude method, finding overall agreement with the latter. However, small differences remain in both the overall magnitude and the positions of the zeros.

Our results for the various FFs are expected to provide useful theoretical input for future studies of the hadronic $\tau$ decays and the semileptonic $D$ meson decays.  

\section*{Acknowledgements}

This work is supported in part by National Natural Science Foundation of China (NSFC) under Grants No.~12475078, No.~12150013, No.~11975090, by the Science Foundation of Hebei Normal University with Contract No.~L2023B09, and the Spanish MCIN/AEI/ 10.13039/501100011033, grant PID2022-136510NB-C31. It has also received funding from the European Union’s Horizon 2020 research and innovation program under grant agreement No.824093. J.R.E. is supported by the Ram\'on y Cajal program RYC2019-027605-I of the Spanish MICIU. JAO acknowledges  partial financial support to the Grant PID2022-136510NB-C32 funded
by \\MCIN/AEI/10.13039/501100011033/ and FEDER, UE. 

\section*{Appendix: Full expressions for form factors in $U(3)$ $\rxt$}

In this appendix, we present the complete expressions for the perturbative FFs, including the strangeness-changing scalar FFs as well as all the vector and tensor FFs. The explicit expressions for the isoscalar and isovector scalar FFs can be found in Ref.~\cite{Guo:2012yt}. It should be particularly noted that, although the original expressions of the FFs are written in terms of the bare decay constant $F$, the final results are re-expressed in terms of the physical pion decay constant $F_\pi$. The relation between the decay constant $F$ in the chiral limit and $F_\pi$ has been explicitly given in Eq.~(C2) of Ref.~\cite{Guo:2011pa}. The higher-order $\eta-\eta'$ mixing matrix elements, $\delta_k$ and $s_\delta=\sin\theta_\delta$, which characterize the mixing between the $\eta$ and $\eta'$ states beyond LO, are explicitly given in Ref.~\cite{Guo:2011pa,Guo:2012yt}, and will not be discussed further here. The explicit expressions of the one-loop functions $A_0(m^2)$ and $B_0(s,m_a^2,m_b^2)$ can also be found in Ref.~\cite{Guo:2011pa}.

The expressions of the strangeness changing scalar FFs read 
\begin{equation}\label{SFFbegin}
\begin{aligned}
F^{\bar{u}s}_{S,K^-\pi^0}&=\frac{F^{\bar{u}s}_{S,\bar{K}^0\pi^-}}{\sqrt{2}}=
\frac{1}{\sqrt{2}}\bigg\{1+\frac{4c_m\big[c_m(m_K^2+m_{\pi}^2)-c_d(m_K^2+m_{\pi}^2-s)\big]}{F_\pi^2(M_{S_8}^2-s)}\\ &-\frac{8(\tilde{c}_d-\tilde{c}_m)\tilde{c}_m(2m_K^2+m_{\pi}^2)}{F_\pi^2M_{S_1}^2}+\frac{4(c_d-c_m)c_m(m_K^2-m_{\pi}^2)}{3F_\pi^2M_{S_8}^2}-\frac{8d_m^2(m_K^2+m_{\pi}^2)}{F_\pi^2M_{P_8}^2}\\ &+\frac{16(m_K^2+m_{\pi}^2)\delta L_8}{F_\pi^2}\bigg\}+\frac{1}{\sqrt{2}}\Bigg\{\frac{(m_K^2-m_{\pi}^2)A_0(m_K^2)}{64F_\pi^2\pi^2s}-\frac{(3m_K^2-3m_{\pi}^2+2s)A_0(m_{\pi}^2)}{128F_\pi^2\pi^2s}\\ &-\frac{\big[c_\theta^2(-9m_K^2+9m_{\pi}^2+6s)+6\sqrt{2}c_\theta(-3m_K^2+3m_{\pi}^2+s)s_\theta-24ss_\theta^2\big]A_0(m_{\eta}^2)}{1152F_\pi^2\pi^2s}\\ &+\frac{\big[8c_\theta^2s+2\sqrt{2}c_\theta(-3m_K^2+3m_{\pi}^2+s)s_\theta+(3m_K^2-3m_{\pi}^2-2s)s_\theta^2\big]A_0(m_{\eta'}^2)}{384F_\pi^2\pi^2s}-\frac{B_0(s,m_{\eta}^2,m_K^2)}{1152F_\pi^2\pi^2s}\bigg\{\\ &c_\theta^2\big[-9m_K^4+(7m_{\pi}^2-9s)s+3m_{\eta}^2(3m_K^2-3m_{\pi}^2+s)+m_K^2(9m_{\pi}^2+2s)\big]-16(2m_K^2+m_{\pi}^2)ss_\theta^2\\ &+2\sqrt{2}c_\theta\bigg[-9m_K^4+m_K^2(9m_{\pi}^2-2s)+(5m_{\pi}^2-9s)s+3m_{\eta}^2(3m_K^2-3m_{\pi}^2+s)\bigg]s_\theta\bigg\}\\ &+\frac{B_0(s,m_{\eta'}^2,m_K^2)}{1152F_\pi^2\pi^2s}\bigg\{16c_\theta^2(2m_K^2+m_{\pi}^2)s+2\sqrt{2}c_\theta\big[-9m_K^4+m_K^2(9m_{\pi}^2-2s)\\ &+(5m_{\pi}^2-9s)s+3m_{\eta'}^2(3m_K^2-3m_{\pi}^2+s)\big]s_\theta +\big[9m_K^4-(7m_{\pi}^2-9s)s-3m_{\eta'}^2(3m_K^2-3m_{\pi}^2+s)\\ &-m_K^2(9m_{\pi}^2+2s)\big]s_\theta^2\bigg\}-\frac{\big[3m_K^4+3m_{\pi}^4+2m_{\pi}^2s-5s^2+m_K^2(-6m_{\pi}^2+2s)\big]B_0(s,m_K^2,m_{\pi}^2)}{128F_\pi^2\pi^2s}\Bigg\}\,,
\end{aligned}
\end{equation}
\begin{equation}
\small
\begin{aligned}
F^{\bar{u}s}_{S,K^-\eta}&=\sqrt{\frac{2}{3}}\bigg\{-\big(\frac{c_\theta}{2}+\sqrt{2}s_\theta \big)+\frac{4(\tilde{c}_d-\tilde{c}_m)\tilde{c}_m(2m_K^2+m_{\pi}^2)(c_\theta+2\sqrt{2}s_\theta)}{F_\pi^2M_{S_1}^2}\\ &+\frac{2c_m\big[c_d(m_{\eta}^2+m_K^2-s)(c_\theta+2\sqrt{2}s_\theta)+c_m(-5c_\theta m_K^2+3c_\theta m_{\pi}^2-4\sqrt{2}m_K^2s_\theta)\big]}{F_\pi^2(M_{S_8}^2-s)}\\ &+\frac{2c_m(m_K^2-m_{\pi}^2)\big[-c_m(11c_\theta+4\sqrt{2}s_\theta)+c_d(c_\theta+2c_\theta^3+2\sqrt{2}s_\theta+8\sqrt{2}c_\theta^2s_\theta+16c_\theta s_\theta^2)\big]}{3F_\pi^2M_{S_8}^2}\\ &+\frac{1}{4}(s_\theta-2\sqrt{2}c_\theta)(\delta_k+2s_\delta)+\frac{1}{4}s_\theta^2(c_\theta+2\sqrt{2}s_\theta)\Lambda_1-\sqrt{2}s_\theta\Lambda_2 \\ & +\frac{4d_m^2(5c_\theta m_K^2-3c_\theta m_{\pi}^2+4\sqrt{2}m_K^2s_\theta)}{F_\pi^2M_{P_8}^2}-\frac{8(5c_\theta m_K^2-3c_\theta m_{\pi}^2+4\sqrt{2}m_K^2s_\theta)\delta L_8}{F_\pi^2}\bigg\}\\ &+\frac{1}{\sqrt{6}}\Bigg\{\frac{\big[c_\theta^3(-9m_{\eta}^2+9m_K^2-10s)-6\sqrt{2}c_\theta^2(3m_{\eta}^2-3m_K^2+4s)s_\theta-24c_\theta ss_\theta^2-16\sqrt{2}ss_\theta^3\big]}{384F_\pi^2\pi^2s}A_0(m_{\eta}^2)\\ &+\frac{\big[-8c_\theta^3s+6\sqrt{2}c_\theta^2(3m_{\eta}^2-3m_K^2+s)s_\theta+3c_\theta(-3m_{\eta}^2+3m_K^2+2s)s_\theta^2-2\sqrt{2}ss_\theta^3\big]}{384F_\pi^2\pi^2s}A_0(m_{\eta'}^2)\\ &+\frac{1}{128F_\pi^2\pi^2s}\bigg\{3c_\theta^3(m_{\eta}^2-m_K^2+s)+c_\theta\bigg[3m_{\eta}^2(-3+s_\theta^2)-3m_K^2(-3+s_\theta^2)-s(3+s_\theta^2)\bigg]\\ &-12\sqrt{2}ss_\theta+8\sqrt{2}c_\theta^2ss_\theta\bigg\}A_0(m_K^2)+\frac{(9c_\theta m_{\eta}^2-9c_\theta m_K^2+6c_\theta s-6\sqrt{2}ss_\theta)}{128F_\pi^2\pi^2s}A_0(m_{\pi}^2)\\ &+\frac{1}{384F_\pi^2\pi^2s}\bigg\{c_\theta^3\bigg[9m_{\eta}^4+9m_K^4-18m_K^2s+4m_{\pi}^2s+9s^2-6m_{\eta}^2(3m_K^2+s)\bigg]\\ &+2\sqrt{2}c_\theta^2\bigg[9m_{\eta}^4+9m_K^4-26m_K^2s+8m_{\pi}^2s+9s^2-6m_{\eta}^2(3m_K^2+s)\bigg]s_\theta+16c_\theta(-5m_K^2+2m_{\pi}^2)ss_\theta^2\\ &-32\sqrt{2}m_K^2ss_\theta^3\bigg\}B_0(s,m_{\eta}^2,m_K^2)+\frac{1}{384F_\pi^2\pi^2s}\bigg\{16c_\theta^3(-2m_K^2+m_{\pi}^2)s\\ &+6\sqrt{2}c_\theta^2\bigg[-3m_K^4+2m_K^2s-2m_{\pi}^2s-3s^2+m_{\eta'}^2(3m_K^2+s)+m_{\eta}^2(-3m_{\eta'}^2+3m_K^2+s)\bigg]s_\theta\\ &+3c_\theta\bigg[3m_K^4+m_{\eta}^2(3m_{\eta'}^2-3m_K^2-s)+10m_K^2s-4m_{\pi}^2s+3s^2-m_{\eta'}^2(3m_K^2+s)\bigg]s_\theta^2\\ &+4\sqrt{2}(-2m_K^2+m_{\pi}^2)ss_\theta^3\bigg\}B_0(s,m_{\eta'}^2,m_K^2)+\frac{1}{128F_\pi^2\pi^2s}\bigg\{-4\sqrt{2}(2m_K^2+m_{\pi}^2)ss_\theta\\ &+c_\theta\bigg[-9m_K^4+(7m_{\pi}^2-9s)s+3m_{\eta}^2(3m_K^2-3m_{\pi}^2+s)+m_K^2(9m_{\pi}^2+2s)\bigg]\bigg\}B_0(s,m_K^2,m_{\pi}^2)\Bigg\}\,,
\end{aligned}
\end{equation}
\begin{equation}\label{SFFend}
\small
\begin{aligned}
F^{\bar{u}s}_{S,K^-\eta'}&=\sqrt{\frac{2}{3}} \Bigg\{ \sqrt{2} c_\theta  - \frac{s_\theta}{2}- \frac{4 (\tilde c_d-\tilde c_m)\tilde c_m (2m_K^2+m_\pi^2) (2\sqrt{2} c_\theta-s_\theta)}{F^2 M_{S_1}^2} + \frac{1}{4} (\delta_k -2s_\delta)( c_\theta + 2\sqrt{2} s_\theta ) \\ & + \frac{2 c_m \big[ c_m \big( 4\sqrt{2} c_\theta m_K^2 - 5 m_K^2 s_\theta + 3 m_\pi^2 s_\theta \big) - c_d (m_{\eta'}^2+m_K^2-s)(2\sqrt{2} c_\theta-s_\theta)  \big]}{F^2 (M_{S_8}^2-s)} \\ &+ \frac{2 c_m (m_K^2-m_\pi^2) \big[ c_m (4\sqrt{2} c_\theta-11 s_\theta) + c_d ( s_\theta + 16 c_\theta^2 s_\theta + 2 s_\theta^3 -  2\sqrt{2} c_\theta - 8\sqrt{2} c_\theta s_\theta^2 ) \big]}{3 F^2 M_{S_8}^2} \\ & + \frac{8 \big( 4\sqrt{2} c_\theta m_K^2 - 5 m_K^2 s_\theta + 3 m_\pi^2 s_\theta \big)\delta L_8}{F^2} - \frac{4 d_m^2 \big( 4\sqrt{2} c_\theta m_K^2 - 5 m_K^2 s_\theta + 3 m_\pi^2 s_\theta \big)}{F^2 M_{P_8}^2} \\ &+ \frac{c_\theta^2 (-2\sqrt{2} c_\theta+s_\theta)\Lambda_1}{4} + \sqrt{2} c_\theta \Lambda_2 \Bigg\}  + \frac{1}{\sqrt{6}} \Bigg\{ \frac{A_0(m_\eta^2)}{384 F^2 \pi^2 s}\bigg[ 2\sqrt{2} c_\theta^3 s - 6\sqrt{2} c_\theta (3m_{\eta'}^2-3m_K^2+s)s_\theta^2 \\ &- 8 s s_\theta^3 + c_\theta^2 \big( -9m_{\eta'}^2 s_\theta + 9m_K^2 s_\theta + 6 s s_\theta \big) \bigg]  + \frac{1}{384 F^2 \pi^2 s}\bigg[ 16\sqrt{2} c_\theta^3 s - 24 c_\theta^2 s s_\theta \\ &+ 6\sqrt{2} c_\theta (3m_{\eta'}^2-3m_K^2+4s)s_\theta^2 + (-9m_{\eta'}^2+9m_K^2-10s)s_\theta^3 \bigg] A_0(m_{\eta'}^2) + \frac{A_0(m_K^2)}{128 F^2 \pi^2 s} \bigg\{ \\ &c_\theta^2 (3m_{\eta'}^2-3m_K^2-s)s_\theta + 4\sqrt{2} c_\theta s (3-2s_\theta^2) + 3 s_\theta \bigg[ m_{\eta'}^2(-3+s_\theta^2) - m_K^2(-3+s_\theta^2) - c_\theta^2 s \bigg] \bigg\}  \\ & + \frac{\big( 6\sqrt{2} c_\theta s + 9m_{\eta'}^2 s_\theta - 9m_K^2 s_\theta + 6 s s_\theta \big) A_0(m_\pi^2)}{128 F^2 \pi^2 s} - \frac{1}{384 F^2 \pi^2 s}\bigg\{ 4\sqrt{2} c_\theta^3 (-2m_K^2+m_\pi^2)s \\ &+ 3 c_\theta^2 \bigg[ -3m_K^4 - 10m_K^2 s + 4m_\pi^2 s - 3s^2 + m_{\eta'}^2 (3m_K^2+s) + m_\eta^2 (-3m_{\eta'}^2+3m_K^2+s) \bigg] s_\theta \\ &+ 6\sqrt{2} c_\theta \bigg[ -3m_K^4 + 2m_K^2 s - 2m_\pi^2 s - 3s^2 + m_{\eta'}^2 (3m_K^2+s) + m_\eta^2 (-3m_{\eta'}^2+3m_K^2+s) \bigg] s_\theta^2 \\ &+ 16 (2m_K^2-m_\pi^2) s s_\theta^3 \bigg\} B_0(s,m_\eta^2,m_K^2) + \frac{ B_0(s,m_{\eta'}^2,m_K^2)}{384 F^2 \pi^2 s} \bigg\{ 16 c_\theta^2 (-5m_K^2+2m_\pi^2) s s_\theta \\ &- 2\sqrt{2} c_\theta \bigg[ 9m_{\eta'}^4 + 9m_K^4 - 26m_K^2 s + 8m_\pi^2 s + 9s^2 - 6m_{\eta'}^2 (3m_K^2+s) \bigg] s_\theta^2 + 32\sqrt{2} c_\theta^3 m_K^2 s \\ &+ \bigg[ 9m_{\eta'}^4 + 9m_K^4 - 18m_K^2 s + 4m_\pi^2 s + 9s^2 - 6m_{\eta'}^2 (3m_K^2+s) \bigg] s_\theta^3 \bigg\}+ \frac{ B_0(s,m_K^2,m_\pi^2)}{128 F^2 \pi^2 s}\bigg\{ \\ & 4\sqrt{2} c_\theta (2m_K^2+m_\pi^2)s + \bigg[ (7m_\pi^2-9s)s - 9m_K^4  + 3m_{\eta'}^2 (3m_K^2-3m_\pi^2+s) + m_K^2 (9m_\pi^2+2s) \bigg] s_\theta \bigg\} \Bigg\}\,. 
\end{aligned}
\end{equation}

The explicit expressions of the various vector FFs are 
\begin{equation}\label{VFFbegin}
\begin{aligned}
F^{\frac{\bar{u}u+\bar{d}d}{\sqrt{2}}}_{+,K^+K^-}=-\frac{1}{\sqrt{2}}\bigg\{1+\frac{F_VG_Vs}{F_\pi^2(M_\omega^2-s)}+\frac{s-6m_K^2}{96F_\pi^2\pi^2}+\frac{A_0(m_K^2)}{16F_\pi^2\pi^2}+\frac{(s-4m_K^2)B_0(s,m_K^2,m_K^2)}{64F_\pi^2\pi^2}\bigg\}\,,
\end{aligned}
\end{equation}
\begin{equation}
\begin{aligned}
F^{\bar{s}s}_{+,K^+K^-}=1+\frac{F_VG_Vs}{F_\pi^2(M_\phi^2-s)}+\frac{s-6m_K^2}{96F_\pi^2\pi^2}+\frac{A_0(m_K^2)}{16F_\pi^2\pi^2}-\frac{(4m_K^2-s)B_0(s,m_K^2,m_K^2)}{64F_\pi^2\pi^2}\,,
\end{aligned}
\end{equation}
\begin{equation}
\begin{aligned}
F^{\bar{u}d}_{+,\pi^-\pi^0}&= -\sqrt{2}\bigg\{1+\frac{F_VG_V\,s}{F_\pi^2\left(M_\rho^2-s\right)}+\frac{s-2m_K^2-4m_{\pi}^2}{96F_\pi^2\pi^2}+\frac{A_0\!\left(m_K^2\right)}{48F_\pi^2\pi^2}\\ &+\frac{A_0\!\left(m_{\pi}^2\right)}{24F_\pi^2\pi^2} +\frac{\left(s-4m_K^2\right)B_0\!\left(s,m_K^2,m_K^2\right)}{192F_\pi^2\pi^2}+\frac{\left(s-4m_{\pi}^2\right)B_0\!\left(s,m_{\pi}^2,m_{\pi}^2\right)}{96F_\pi^2\pi^2}\bigg\}\,,
\end{aligned}
\end{equation}

\begin{equation}
\begin{aligned}
F^{\bar{u}d}_{+,K^0K^-}&=-\bigg\{1+\frac{F_VG_V\,s}{F_\pi^2\left(M_\rho^2-s\right)}+\frac{s-2m_K^2-4m_{\pi}^2}{96F_\pi^2\pi^2}+\frac{A_0\!\left(m_K^2\right)}{48F_\pi^2\pi^2}\\ &+\frac{A_0\!\left(m_{\pi}^2\right)}{24F_\pi^2\pi^2}+\frac{\left(s-4m_K^2\right)B_0\!\left(s,m_K^2,m_K^2\right)}{192F_\pi^2\pi^2}+\frac{\left(s-4m_{\pi}^2\right)B_0\!\left(s,m_{\pi}^2,m_{\pi}^2\right)}{96F_\pi^2\pi^2}\bigg\}\,,
\end{aligned}
\end{equation}
\begin{equation}
\begin{aligned}
F^{\bar{u}s}_{+,K^-\pi^0}&=\frac{1}{\sqrt{2}} F^{\bar{u}s}_{+,\bar{K}^0\pi^-}=-\frac{1}{\sqrt{2}}\bigg\{1+\frac{F_VG_Vs}{F_\pi^2(M_{K^*}^2-s)}\bigg\}+\frac{1}{\sqrt{2}}\bigg\{\frac{3m_K^2+3m_{\pi}^2-s+c_\theta^2(3m_{\eta}^2+3m_K^2-s)}{192F_\pi^2\pi^2}\\ &+\frac{(3m_{\eta'}^2+3m_K^2-s)s_\theta^2}{192F_\pi^2\pi^2}+\frac{c_\theta^2(m_{\eta}^2-m_K^2-2s)A_0(m_{\eta}^2)}{128F_\pi^2\pi^2 s}+\frac{(m_{\eta'}^2-m_K^2-2s)s_\theta^2A_0(m_{\eta'}^2)}{128F_\pi^2\pi^2 s}\\ &+\bigg[\frac{m_K^2-m_{\pi}^2-5s+c_\theta^2(-m_{\eta}^2+m_K^2+s)+(-m_{\eta'}^2+m_K^2+s)s_\theta^2}{128F_\pi^2\pi^2 s}\bigg]A_0(m_K^2)\\ &-\frac{(m_K^2-m_{\pi}^2+2s)A_0(m_{\pi}^2)}{128F_\pi^2\pi^2 s}-\frac{c_\theta^2\big[m_{\eta}^4+(m_K^2-s)^2-2m_{\eta}^2(m_K^2+s)\big]B_0(s,m_{\eta}^2,m_K^2)}{128F_\pi^2\pi^2 s}\\ &-\frac{s_\theta^2\big[m_{\eta'}^4+(m_K^2-s)^2-2m_{\eta'}^2(m_K^2+s)\big] B_0(s,m_{\eta'}^2,m_K^2)}{128F_\pi^2\pi^2 s}\\ &-\frac{\big[m_K^4+(m_{\pi}^2-s)^2-2m_K^2(m_{\pi}^2+s)\big]B_0(s,m_K^2,m_{\pi}^2)}{128F_\pi^2\pi^2 s}\bigg\}\,,
\end{aligned}
\end{equation}
\begin{equation}
\begin{aligned}
F^{\bar{u}s}_{+,K^-\eta}&=-\sqrt{\frac{3}{2}}c_\theta\bigg\{1+\frac{ F_VG_Vs}{F_\pi^2(M_{K^*}^2-s)}\bigg\}+\frac{4c_dc_m(m_K^2-m_{\pi}^2)(-\sqrt{2}c_\theta+\sqrt{2}c_\theta^3-2s_\theta+4c_\theta^2s_\theta)}{\sqrt{3}F_\pi^2M_{S_8}^2}\\ &+\frac{1}{2}\sqrt{\frac{3}{2}}\bigg\{s_\theta(2s_\delta+\delta_k)+c_\theta s_\theta^2\Lambda_1\bigg\} + \sqrt{\frac{3}{2}}\bigg\{\frac{c_\theta\big[3m_K^2+3m_{\pi}^2+c_\theta^2(3m_{\eta}^2+3m_K^2-s)-s\big]}{192F_\pi^2\pi^2}\\ & +\frac{c_\theta(3m_{\eta'}^2+3m_K^2-s)s_\theta^2}{192 F_\pi^2\pi^2}+\frac{c_\theta^3(m_{\eta}^2-m_K^2-2s)A_0(m_{\eta}^2)}{128F_\pi^2\pi^2 s}+\frac{c_\theta(m_{\eta'}^2-m_K^2-2s)s_\theta^2A_0(m_{\eta'}^2)}{128F_\pi^2\pi^2 s}\\ &+\frac{c_\theta\big[m_K^2-m_{\pi}^2-9s+c_\theta^2(-m_{\eta}^2+m_K^2+5s)\big]+c_\theta(-m_{\eta'}^2+m_K^2+s)s_\theta^2}{128F_\pi^2\pi^2 s}A_0(m_K^2)\\ &-\frac{c_\theta(m_K^2-m_{\pi}^2+2s)A_0(m_{\pi}^2)}{128F_\pi^2\pi^2 s}-\frac{c_\theta^3\big[m_{\eta}^4+(m_K^2-s)^2-2m_{\eta}^2(m_K^2+s)\big]B_0(s,m_{\eta}^2,m_K^2)}{128F_\pi^2\pi^2 s}\\ &-\frac{c_\theta s_\theta^2\big[m_{\eta'}^4+(m_K^2-s)^2-2m_{\eta'}^2(m_K^2+s)\big]B_0(s,m_{\eta'}^2,m_K^2)}{128F_\pi^2\pi^2 s}\\ &-\frac{c_\theta\big[m_K^4+(m_{\pi}^2-s)^2-2m_K^2(m_{\pi}^2+s)\big]B_0(s,m_K^2,m_{\pi}^2)}{128F_\pi^2\pi^2 s}\bigg\}\,,
\end{aligned}
\end{equation}
\begin{equation}\label{VFFend}
\begin{aligned}
F^{\bar{u}s}_{+,K^-\eta'}&=-\sqrt{\frac{3}{2}}s_\theta\bigg\{1+\frac{F_VG_Vs}{F_\pi^2(M_{K^*}^2-s)}\bigg\}+\frac{4c_dc_m(m_K^2-m_{\pi}^2)\big[c_\theta(2-4s_\theta^2)+\sqrt{2}s_\theta(-1+s_\theta^2)\big]}{\sqrt{3}F_\pi^2M_{S_8}^2}\\ &+\frac{1}{2}\sqrt{\frac{3}{2}}\bigg\{c_\theta(-2s_\delta+\delta_k)+c_\theta^2s_\theta\Lambda_1\bigg\}+\frac{1}{\sqrt{2}}\Bigg\{\frac{\big[3m_K^2+3m_{\pi}^2+c_\theta^2(3m_{\eta}^2+3m_K^2-s)-s\big]s_\theta}{64\sqrt{3}F_\pi^2\pi^2}\\ & +\frac{(3m_{\eta'}^2+3m_K^2-s)s_\theta^3}{64\sqrt{3}F_\pi^2\pi^2}+\frac{\sqrt{3}c_\theta^2(m_{\eta}^2-m_K^2-2s)s_\theta A_0(m_{\eta}^2)}{128F_\pi^2\pi^2 s}+\frac{\sqrt{3}(m_{\eta'}^2-m_K^2-2s)s_\theta^3A_0(m_{\eta'}^2)}{128F_\pi^2\pi^2 s}\\ &+\bigg\{\frac{\sqrt{3}\big[m_K^2-m_{\pi}^2-9s+c_\theta^2(-m_{\eta}^2+m_K^2+s)\big]s_\theta}{128F_\pi^2\pi^2 s}+\frac{\sqrt{3}(-m_{\eta'}^2+m_K^2+5s)s_\theta^3}{128F_\pi^2\pi^2 s}\bigg\}A_0(m_K^2)\\ &-\frac{\sqrt{3}(m_K^2-m_{\pi}^2+2s)s_\theta A_0(m_{\pi}^2)}{128F_\pi^2\pi^2 s}-\frac{\sqrt{3}c_\theta^2\big[m_{\eta}^4+(m_K^2-s)^2-2m_{\eta}^2(m_K^2+s)\big]s_\theta B_0(s,m_{\eta}^2,m_K^2)}{128F_\pi^2\pi^2 s}\\ &-\frac{\sqrt{3}\big[m_{\eta'}^4+(m_K^2-s)^2-2m_{\eta'}^2(m_K^2+s)\big]s_\theta^3B_0(s,m_{\eta'}^2,m_K^2)}{128F_\pi^2\pi^2 s}\\ &-\frac{\sqrt{3}\big[m_K^4+(m_{\pi}^2-s)^2-2m_K^2(m_{\pi}^2+s)\big]s_\theta B_0(s,m_K^2,m_{\pi}^2)}{128F_\pi^2\pi^2 s}\Bigg\}\,.
\end{aligned}
\end{equation}

The explicit expressions of the tensor FFs are given by 
\begin{equation}\label{TFFbegin}
\begin{aligned}
F^{\frac{\bar{u}u+\bar{d}d}{\sqrt{2}}}_{T,K^+K^-}&=-\frac{1}{\sqrt{2}}\bigg\{1+\frac{\sqrt{2}F_{V}^TG_VM_V}{\Lambda_2^T(M_\omega^2-s)}+\frac{8c_dc_m(m_{\pi}^2-m_K^2)}{F_\pi^2M_{S_8}^2} + \frac{s-6m_K^2}{96F_\pi^2\pi^2}+\frac{(c_\theta^2-2\sqrt{2}c_\theta s_\theta+2s_\theta^2)A_0(m_{\eta}^2)}{96F_\pi^2\pi^2}\\ &+\frac{(2c_\theta^2+2\sqrt{2}c_\theta s_\theta+s_\theta^2)A_0(m_{\eta'}^2)}{96F_\pi^2\pi^2}+\frac{7A_0(m_{\pi}^2)}{32F_\pi^2\pi^2}+\frac{(s-4m_K^2)B_0(s,m_K^2,m_K^2)}{64F_\pi^2\pi^2}\bigg\}\,,
\end{aligned}
\end{equation}
\begin{equation}
\begin{aligned}
F^{\bar{s}s}_{T,K^+K^-}&=1+\frac{\sqrt2 F_{V}^TG_VM_V}{\Lambda_2^T(M_\phi^2-s)}-\frac{8c_dc_m(m_K^2-m_{\pi}^2)}{F_\pi^2M_{S_8}^2}-\bigg\{\frac{6m_K^2-s}{96F_\pi^2\pi^2}\\ &-\frac{(2c_\theta^2+2\sqrt2c_\theta s_\theta+s_\theta^2)A_0(m_{\eta}^2)}{48F_\pi^2\pi^2}-\frac{(c_\theta^2-2\sqrt2c_\theta s_\theta+2s_\theta^2)A_0(m_{\eta'}^2)}{48F_\pi^2\pi^2}\\ &-\frac{A_0(m_K^2)}{16F_\pi^2\pi^2}-\frac{A_0(m_{\pi}^2)}{8F_\pi^2\pi^2}+\frac{(4m_K^2-s)B_0(s,m_K^2,m_K^2)}{64F_\pi^2\pi^2}\bigg\}\,,
\end{aligned}
\end{equation}
\begin{equation}
\begin{aligned}
F^{\bar{u}d}_{T,\pi^-\pi^0}&= -\sqrt{2}\bigg\{1+\frac{\sqrt2 F_V^TG_VM_V}{\Lambda_2^T(M_\rho^2-s)}-\frac{2m_K^2+4m_{\pi}^2-s}{96 F_\pi^2\pi^2}+\frac{(c_\theta^2-2\sqrt{2}c_\theta s_\theta+2s_\theta^2)A_0(m_{\eta}^2)}{96F_\pi^2\pi^2}\\ &+\frac{(2c_\theta^2+2\sqrt{2}c_\theta s_\theta+s_\theta^2)A_0(m_{\eta'}^2)}{96F_\pi^2\pi^2}+ \frac{A_0(m_K^2)}{12F_\pi^2\pi^2}+\frac{13A_0(m_{\pi}^2)}{96F_\pi^2\pi^2}\\ &+\frac{(s-4m_K^2)B_0(s,m_K^2,m_K^2)}{192F_\pi^2\pi^2}+\frac{(s-4m_{\pi}^2)B_0(s,m_{\pi}^2,m_{\pi}^2)}{96F_\pi^2\pi^2}\bigg\}\,,
\end{aligned}
\end{equation}
\begin{equation}
\begin{aligned}
F^{\bar{u}d}_{T,K^0K^-}&= -\bigg\{1+\frac{\sqrt{2}F_V^TG_VM_V}{\Lambda_2^T(M_\rho^2-s)}-\frac{8c_dc_m(m_K^2-m_{\pi}^2)}{F_\pi^2M_{S_8}^2} + \frac{s-2m_K^2-4m_{\pi}^2}{96F_\pi^2\pi^2}\\ &+\frac{(\sqrt{2}c_\theta^2-4c_\theta s_\theta+2\sqrt{2}s_\theta^2)A_0(m_{\eta}^2)}{96\sqrt2 F_\pi^2\pi^2}+\frac{(2\sqrt{2}c_\theta^2+4c_\theta s_\theta+\sqrt{2}s_\theta^2)A_0(m_{\eta'}^2)}{96\sqrt2F_\pi^2\pi^2}+\frac{A_0(m_K^2)}{12F_\pi^2\pi^2}\\ &+\frac{13A_0(m_{\pi}^2)}{96 F_\pi^2\pi^2}+\frac{(-4m_K^2+s)B_0(s,m_K^2,m_K^2)}{192 F_\pi^2\pi^2}+\frac{(-4m_{\pi}^2+s)B_0(s,m_{\pi}^2,m_{\pi}^2)}{96F_\pi^2\pi^2}\bigg\}\,,
\end{aligned}
\end{equation}
\begin{equation}
\begin{aligned}
F^{\bar{u}s}_{T,K^-\pi^0}&=\frac{1}{\sqrt{2}} F^{\bar{u}s}_{T,\bar{K}^0\pi^-}=-\frac{1}{\sqrt{2}}\bigg\{1+\frac{\sqrt{2}F_V^TG_VM_V}{\Lambda_2^T(M_{K^*}^2-s)}-\frac{4c_dc_m(m_K^2-m_{\pi}^2)}{F_\pi^2M_{S_8}^2}\bigg\}\\ &-\frac{-3m_{\pi}^2+s+c_\theta^2(-3m_{\eta}^2-3m_K^2+s)-3m_{\eta'}^2 s_\theta^2+s s_\theta^2-3m_K^2(1+s_\theta^2)}{192\sqrt2F_\pi^2\pi^2 }\\ &+\frac{1}{\sqrt{2}}\bigg\{-\frac{\big[ c_\theta^2(-3m_{\eta}^2+3m_K^2-2s)+4\sqrt{2}c_\theta s s_\theta+8s s_\theta^2\big]A_0(m_{\eta}^2)}{384F_\pi^2\pi^2 s}\\ &+\frac{\big[-8c_\theta^2 s+4\sqrt{2}c_\theta s s_\theta+(3m_{\eta'}^2-3m_K^2+2s)s_\theta^2\big]A_0(m_{\eta'}^2)}{384F_\pi^2\pi^2 s}\\ &+\frac{\big[-m_{\pi}^2-13s+c_\theta^2(-m_{\eta}^2+m_K^2+s)-m_{\eta'}^2 s_\theta^2+s s_\theta^2+m_K^2(1+s_\theta^2)\big]A_0(m_K^2)}{128F_\pi^2\pi^2 s}\\ &-\frac{3(m_K^2-m_{\pi}^2+18s)A_0(m_{\pi}^2)}{384F_\pi^2\pi^2 s}-\frac{c_\theta^2\big[m_{\eta}^4+(m_K^2-s)^2-2m_{\eta}^2(m_K^2+s)\big]B_0(s,m_{\eta}^2,m_K^2)}{128F_\pi^2\pi^2 s}\\ &-\frac{\big[m_{\eta'}^4+(m_K^2-s)^2-2m_{\eta'}^2(m_K^2+s)\big]s_\theta^2B_0(s,m_{\eta'}^2,m_K^2)}{128F_\pi^2\pi^2 s}\\ &-\frac{\big[m_K^4+(m_{\pi}^2-s)^2-2m_K^2(m_{\pi}^2+s)\big]B_0(s,m_K^2,m_{\pi}^2)}{128F_\pi^2\pi^2 s}\bigg\}\,,
\end{aligned}
\end{equation}
\begin{equation}
\begin{aligned}
F^{\bar{u}s}_{T,K^-\eta}&=-\sqrt{\frac{3}{2}}c_\theta\bigg\{1+\frac{\sqrt{2} F_V^TG_VM_V}{\Lambda_2^T(M_{K^*}^2-s)}-\frac{4c_dc_m(m_K^2-m_{\pi}^2)(5+2c_\theta^2+4\sqrt{2}c_\theta s_\theta)}{3F_\pi^2M_{S_8}^2}\bigg\}\\ &+\frac{1}{2}\sqrt{\frac{3}{2}}\bigg\{s_\theta(\delta_k+2s_\delta)+c_\theta s_\theta^2\Lambda_1\bigg\}\\ &+\sqrt{\frac{3}{2}}\bigg\{-\frac{c_\theta\big[c_\theta^2(3m_K^2-3m_{\eta}^2-2s)+4\sqrt{2}c_\theta s_\theta s +8 s_\theta^2 s \big]A_0(m_{\eta}^2)}{384F_\pi^2\pi^2 s}\\ &-\frac{c_\theta\big[8c_\theta^2 s-4\sqrt{2}c_\theta s s_\theta+(-3m_{\eta'}^2+3m_K^2-2s)s_\theta^2\big]A_0(m_{\eta'}^2)}{384F_\pi^2\pi^2 s}-\frac{c_\theta(m_K^2-m_{\pi}^2+18s)A_0(m_{\pi}^2)}{128F_\pi^2\pi^2 s}\\ &+\frac{c_\theta\big[-m_{\pi}^2-17s+c_\theta^2(-m_{\eta}^2+m_K^2+5s)-m_{\eta'}^2 s_\theta^2+s s_\theta^2+m_K^2(1+s_\theta^2)\big]A_0(m_K^2)}{128F_\pi^2\pi^2 s}\\ &-\frac{c_\theta\big[-3m_{\pi}^2+s+c_\theta^2(-3m_{\eta}^2-3m_K^2+s)-3m_{\eta'}^2 s_\theta^2+s s_\theta^2-3m_K^2(1+s_\theta^2)\big]}{192F_\pi^2\pi^2 }\\ &-\frac{c_\theta^3\big[m_{\eta}^4+(m_K^2-s)^2-2m_{\eta}^2(m_K^2+s)\big]B_0(s,m_{\eta}^2,m_K^2)}{128F_\pi^2\pi^2 s}\\ &-\frac{c_\theta\big[m_{\eta'}^4+(m_K^2-s)^2-2m_{\eta'}^2(m_K^2+s)\big]s_\theta^2B_0(s,m_{\eta'}^2,m_K^2)}{128F_\pi^2\pi^2 s}\\ &-\frac{c_\theta\big[m_K^4+(m_{\pi}^2-s)^2-2m_K^2(m_{\pi}^2+s)\big]B_0(s,m_K^2,m_{\pi}^2)}{128F_\pi^2\pi^2 s}\bigg\}\,,
\end{aligned}
\end{equation}
\begin{equation}\label{TFFend}
\begin{aligned}
F^{\bar{u}s}_{T,K^-\eta'}&=-\sqrt{\frac{3}{2}}s_\theta\bigg\{1+\frac{\sqrt{2}F_V^TG_VM_V}{\Lambda_2^T(M_{K^*}^2-s)}-\frac{4c_dc_m(m_K^2-m_{\pi}^2)(5-4\sqrt{2}c_\theta s_\theta+2s_\theta^2)}{3F_\pi^2M_{S_8}^2}\bigg\}\\ &+\frac{1}{2}\sqrt{\frac{3}{2}}\bigg\{c_\theta(\delta_k-2s_\delta)+c_\theta^2s_\theta\Lambda_1\bigg\}\\ &+\sqrt{\frac{3}{2}}\bigg\{-\frac{s_\theta\big[c_\theta^2(-3m_{\eta}^2+3m_K^2-2s)+4\sqrt{2}c_\theta s_\theta s +8s s_\theta^2\big]A_0(m_{\eta}^2)}{384F_\pi^2\pi^2s}\\ &+\frac{s_\theta\big[-8c_\theta^2s+4\sqrt{2}c_\theta s_\theta s +(3m_{\eta'}^2-3m_K^2+2s)s_\theta^2\big]A_0(m_{\eta'}^2)}{384F_\pi^2\pi^2s}-\frac{s_\theta(m_K^2-m_{\pi}^2+18s)A_0(m_{\pi}^2)}{128 F_\pi^2\pi^2s}\\ &+\frac{s_\theta\big[-m_{\pi}^2-17s+c_\theta^2(-m_{\eta}^2+m_K^2+s)-m_{\eta'}^2s_\theta^2+5ss_\theta^2+m_K^2(1+s_\theta^2)\big]A_0(m_K^2)}{128F_\pi^2\pi^2s}\\ &-\frac{2s s_\theta\big[-3m_{\pi}^2+s+c_\theta^2(-3m_{\eta}^2-3m_K^2+s)-3m_{\eta'}^2s_\theta^2+ss_\theta^2-3m_K^2(1+s_\theta^2)\big]}{384F_\pi^2\pi^2s}\\ &-\frac{c_\theta^2\big[m_{\eta}^4+(m_K^2-s)^2-2m_{\eta}^2(m_K^2+s)\big]s_\theta B_0(s,m_{\eta}^2,m_K^2)}{128F_\pi^2\pi^2s}\\ &-\frac{\big[m_{\eta'}^4+(m_K^2-s)^2-2m_{\eta'}^2(m_K^2+s)\big]s_\theta^3B_0(s,m_{\eta'}^2,m_K^2)}{128F_\pi^2\pi^2s}\\ &-\frac{\big[m_K^4+(m_{\pi}^2-s)^2-2m_K^2(m_{\pi}^2+s)\big]s_\theta B_0(s,m_K^2,m_{\pi}^2)}{128F_\pi^2\pi^2s}\bigg\}\,.
\end{aligned}
\end{equation}

\bibliographystyle{apsrev4-2}
\bibliography{u3ff}

\end{document}